\documentclass[structabstract]{aa}  
\usepackage{graphicx}
\usepackage{float}
\usepackage{txfonts}
\usepackage{natbib}
\bibpunct{(}{)}{;}{a}{}{,} 
\usepackage{colortbl}
\usepackage{amssymb}
\usepackage{epsf}
\usepackage{natbib}
\usepackage{rotating}
\usepackage{booktabs}
\usepackage{subfigure}
\usepackage{enumerate}
\usepackage{color}
\usepackage{amssymb}
\usepackage{pifont}

\usepackage{caption}

\definecolor{light-gray}{gray}{0.8}
\begin{document}

   \title{Impact of stellar companions on precise radial velocities}


   \author{D. Cunha
          \inst{1,2}
          \and
          P. Figueira \inst{1}
          \and
          N. C. Santos \inst{1,2}
          \and
           C. Lovis\inst{3}
          \and
          G. Bou\'e\inst{1,4}}

   \institute{Centro de Astrof\'{i}sica da Universidade do Porto, Rua das Estrelas, 4150-762 Porto, Portugal\\
              \email{Diana.Cunha@astro.up.pt}
         \and
             Departamento de F\'{i}sica e Astronomia, Faculdade de Ci\^encias, Universidade do Porto, 
	     Rua do Campo Alegre 687, 4169-007 Porto, Portugal
         \and
            Observatoire Astronomique de l'Universit\'e de Gen\`eve,
            51 Ch. des Maillettes, - Sauverny - CH1290, Versoix, Suisse
         \and
          ASD, IMCCE-CNRS UMR8028, Observatoire de Paris, UPMC, 77 avenue Denfert-Rochereau,
          75014 Paris, France}
    \date{Received \today; accepted November 29, 2012}

 
  \abstract
    {With the announced arrival of instruments such as ESPRESSO one can  expect that several systematic noise sources on the measurement 
of precise radial velocity will become the limiting factor 
instead of photon noise.
A stellar companion within the fiber is such a possible noise source.}
    {With this work we aim at characterizing the impact of  a stellar companion within the fiber to radial velocity measurements made by fiber-fed
spectrographs.
We consider the contaminant star either to be part of a binary system whose primary star is the target star, or  
as a background/foreground star.}
    {To carry out our study, we used HARPS spectra, co-added the target with contaminant spectra, and then compared the resulting radial
 velocity with that obtained 
from the original target spectrum. We repeated this procedure and used different tunable knobs to reproduce the previously mentioned scenarios.
}
{We find that the impact on the radial velocity  calculation 
is a function of the difference between individual radial
velocities, of the difference between target and contaminant magnitude, and also of their spectral types.
 For the worst-case scenario in which both target and contaminant star {are well centered on the fiber},
the maximum contamination for a G or K star  may be higher than 10 cm/s, on average,
 if the difference between target and 
contaminant magnitude is  $\Delta m < 10$, and higher than 1 m/s if $\Delta m < 8$. 
If the target star is of spectral type M, 
  $\Delta m < 8$ produces the same contamination of 10 cm/s, and a contamination may be higher than 1 m/s  if $\Delta m < 6$. }
   {}
\maketitle
%

\section{Introduction}
{The search for extrasolar planets is currently a very active field of research in astronomy. 
Since the first discovery of \citet{Mayor_Queloz_1995}, more than 800 planets were discovered 
\footnote{ http://exoplanet.eu/, as of 27/09/2012}, of which 80 \% were detected with the 
radial velocity method (RV). Because it is the workhorse for planetary detections, this method received 
much attention from the community, which constantly increased its precision \citep[for some of the latest results
 check][]{Mayor_et_al_2011} and tried to characterize its limitations.

One of the fundamental drawbacks is that it is an indirect method, and therefore one should be extremely 
careful with false positives, i.e., signals created by other planetary RV signals. 
The literature provides many examples and techniques used to pinpoint these signals 
\citep[e.g.][]{Queloz_et_al_2001, Santos_2002, Melo_2007,Huelamo_et_al_2008, Figueira_2010} or to average them out
 \citep{Dumusque_2011a, Dumusque_2011b}. With the advent of very high precision spectrographs, such as ESPRESSO or CODEX 
\citep{Pepe_2010}, the RV precision enters a new domain, that of cm/s,
 which in turn will increase the quality of our characterization of these unwanted yet ubiquitous signals \citep[e.g.][]{Cegla_2012}.

Here we address a so far unexplored mechanism that is capable of distorting RV signals and
 creating false ones: the contamination of the main spectrum by that of a (usually unresolved or otherwise undetectable)
 companion. 
We consider two different cases: 
\begin{enumerate}
\item that of a faint gravitationally bounded companion and
\item that of an unbound star that is aligned with the target at the moment of observation. 
\end{enumerate} 
We assumed the target stars to be placed at the Galaxy disk edge to consider a representative case that 
encompasses several scenarios of contaminant stars.
We used 
real high signal to noise ratio (S/N) observations obtained with HARPS to create the composite spectra and process these with 
the RV calculation pipeline  to evaluate the impact of the contamination in the most realistic way possible.
While we aim at characterizing this effect down to the precision level of ESPRESSO, 10 cm/s, it can be assessed 
at an even lower precision level. 

Although this study is performed to characterize the impact on the RV when using the cross-correlation function (CCF), 
point spread function (PSF) modeling will also be affected by stellar companions, if in a different way.
The results of this study are particularly interesting for deep surveys in crowded fields (Kepler).

If the contaminant RV remains constant, stellar blends will only cause a undetectable  contamination in the RV calculation. 
However, if the contaminant RV is changing, if the seeing varies, or if there are guiding problems during observations, the effect of the 
contamination will no longer be constant and it might mimic the presence of a planet. 
As an example of a false positive caused by blended stars we refer to the work of \cite{Santos_2002},
 in which RV measurements derived from CORALIE blended spectra of HD 41004 AB have unveiled a
radial-velocity variation with a period of 1.3 days and a small
amplitude of 50 m/s, compatible with the signal expected
to be caused by a planetary companion to HD 41004 A. Another example is WASP-9b,  which was 
discarded as a planet after it was discovered that the signal was due to a fortuitous alignment (Triaud, priv. comm.).

 In Sect. \ref{sec:CC} we explain which stars we consider most likely to contribute to a shift in the RV calculation,
 if the star is either in a G-, K- or M-dwarf binary system, or  is affected by a fortuitous alignment. 
Then, in Sect. \ref{sec:data} we describe the properties of the spectra used in our simulations.
In Sect. \ref{sec:Met} we describe the method used to scramble spectra 
from target star and stellar companion.  
Results from our work, assuming the worst-case scenario in which both {target and contaminant are well centered on the fiber,}
  are shown in  Sect. \ref{sec:Resul}.
 The results are followed by a statistical analysis in  Sect.  \ref{sec:SA},
 which is based on these results, but assumes a more realistic situation, in which the contaminant star may not be 
fiber centered. In  Sect. \ref{sec:Disc} we  discuss our work, and we 
  present our conclusions in  Sect. \ref{sec:Concl}.   
}
  
\section{Contaminant cases}\label{sec:CC}
{}
{When  one observes a target star, a secondary star may be present, even without one's knowledge.
 This secondary star may be a source of contamination for the calculation of the target star's RV. 
Secondary/contaminants stars may appear as two different types: Real binaries or
 fortuitous alignments.} 

\subsection{Real binaries}\label{Bin1}
In the case of a real binary system, the contaminant star will be gravitationally bound. 
In this section we present the probability distributions of real binary properties that we used in our work.

Here, for G- and K primaries we based our study 
on the results  
 of \cite{Duquennoy_Mayor_1991}, who studied the multiplicity among solar-type stars
 in the solar neighborhood. 
They considered a subsample of 164 primary stars with spectral types between F7 and G9.
 Using  Table 7 of \cite{Duquennoy_Mayor_1991} on the mass-ratio distribution among G-dwarfs binaries, 
we can calculate the probability 
of observing a binary with a mass-ratio $q=M_2/M_1$, where $M_2$ is the mass of the secondary star and $M_1$ 
the mass of the primary.  
These probabilities are shown in Table \ref{tab:P|B}, in which we can see that there is a maximum for the 
mass-ratio range ]0.2,0.3]. The distribution of orbital periods (P)
of their study is approximated to a Gaussian with $\overline{\log P} =4.8$ and $\sigma_{log P} = 2.3$. 

Duquennoy \& Mayor also presented the eccentricity distribution for  $P < 1000$ and $P > 1000$ days,
shown here  in Table \ref{tab:Pecc}.

We stress that there is a large uncertainty for  VLMC (very low mass companion) binaries (with $q<0.1$), 
and that this study does not include values of $q$ below 0.01. We also emphasize that this mass ratio
 distribution 
is calculated for binary stars, but we recall that, also according to \cite{Duquennoy_Mayor_1991}, 
there are 57\% of binary systems
with a mass ratio higher than 0.1, and 43\% of apparently single stars. Of these 43\%, ($8\pm6$)\%  
 most probably  have a VLMC, 
which puts the percentage of  single stars at $\sim 30$\%. 

\begin {table}[]
 \caption{ Probability of observing a binary with a mass ratio $q=M_2/M_1$ among G-dwarf binaries for periods $P<10^4$ days and $P<10^4$ 
\citep[from][]{Duquennoy_Mayor_1991}.}
 \begin{center}
  \tabcolsep 2.8pt
 \small
 \begin{tabular}{c| c c c c c c c c c c c}
   \hline
 \textbf{q$_{max}$} &  0.1  &  0.2  &   0.3  &   0.4 &   0.5 &   0.6 &  0.7  &   0.8 &   0.9 &  1.0 & 1.1 \\\hline
&\multicolumn{11}{c}{$P< 10^4 days$} \\
\textbf{P(q) [\%]}& 12.0 & 16.3 & 13.1 & 12.7 & 12.3 & 12.9 & 6.5 & 1.9 & 2.6 & 3.4 & 6.2 \\\hline
&\multicolumn{11}{c}{$P> 10^4 days$} \\
\textbf{ P(q) [\%]}& 11.1 & 11.9 & 18.4 & 15.8 & 10.5 & 11.2 & 6.6 & 6.6 & 4.0 & 4.0 &  0 \\\hline
&\multicolumn{11}{c}{Total} \\
\textbf{ P(q) [\%]}    & 11.4 & 13.6 & 16.4 & 14.6 & 11.2 & 11.8 & 6.5 & 4.8 & 3.4 & 3.8 & 2.4 \\\hline
 \end{tabular}

 \end{center}
 \label{tab:P|B}
 \end{table}

\begin{table}[]
 \caption{ Probability of observing a binary with an eccentricity $e$ among G-dwarfs binaries for periods $P<10^3$ days and $P<10^3$ 
\citep[from][]{Duquennoy_Mayor_1991}.}
 \begin{center}
  \tabcolsep 2.8pt
 \small
 \begin{tabular}{c| c c c c c c c}
   \hline
 \textbf{e$_{max}$} &  0.15  &  0.3  &   0.45  &   0.6 &   0.75 &   0.9 &  1   \\\hline
 &\multicolumn{7}{c}{$P< 10^3 days$} \\
\textbf{P(e) [\%]}& 12.50 & 43.75 & 31.25 & 6.25 & 6.25 & 0.0 & 0.0  \\\hline
 &\multicolumn{7}{c}{$P> 10^3 days$} \\
\textbf{ P(e) [\%]}& 5.88 & 11.76& 23.53 & 14.71 & 20.59 & 23.53 & 0.0  \\\hline
 \end{tabular}
 \end{center}
 \label{tab:Pecc}
 \end{table}

For M-dwarfs we used the distributions of mass ratio, $q$, and semi-major axis, $a$, from the work of \citet{Janson_2012} on 
M-dwarf multiplicity. The authors suggest an M-dwarf total multiplicity fraction of 34\% .
Table \ref{tab:PqM} shows the probability of observing an M-dwarf binary with a mass ratio $q$.
Janson et al. found a uniform distribution to be more consistent with the mass ratio of their observed sample, than a rising 
distribution. Accordingly, we assumed their uniform distribution in our work. 
For the semi-major axis distribution \citet{Janson_2012} also compared the Sun-like and the narrow distribution, finding the latter to be
more consistent with their sample. We therefore
used this distribution, which is shown in Table \ref{tab:Pa}. As no information was given on the eccentricity 
distribution for M-dwarf binaries, we assumed the distribution from \citet{Duquennoy_Mayor_1991}.

 These distributions were used in our statistical analysis (Sect. \ref{sec:SA}). 

\begin {table}[]
 \caption{ Probability of observing an M-dwarf binary with a mass ratio $q=M_2/M_1$ 
\citep[from][]{Janson_2012}.}
 \begin{center}
  \tabcolsep 2.8pt
 \tiny
 \begin{tabular}{c| c c c c c c c c c c }
   \hline
 \textbf{q$_{max}$} &  0.1  &  0.2  &   0.3  &   0.4 &   0.5 &   0.6 &  0.7  &   0.8 &   0.9 &  1.0  \\\hline
\textbf{P(q) [\%]}  &  0.0  &  1.61 &  5.95  & 10.45 & 12.93 & 13.84 & 14.48 & 13.03 & 14.58 & 13.13   \\\hline

 \end{tabular}

 \end{center}
 \label{tab:PqM}
 \end{table}

\begin {table}[]
 \caption{ Probability of observing an M-dwarf binary with a certain $\log a$  
\citep[from][]{Janson_2012}.}
 \begin{center}
  \tabcolsep 2.8pt
 \tiny
 \begin{tabular}{c| c c c c c c c c c c }
   \hline
 \textbf{\textbf{log}$a_{max}$} &  0.25  &  0.5  &   0.75  &   1.  &   1.25 &   1.5 &  1.75  &   2.  &   2.25 &  2.5  \\\hline
\textbf{P({log}$a$) [\%]}  &  0.51  &  2.60 &   9.02  & 15.39 &  17.13 & 17.65 & 14.73  & 11.74 &   7.85 & 3.37   \\\hline

 \end{tabular}
 \end{center}
 \label{tab:Pa}
 \end{table}

\subsection{Fortuitous alignment}\label{AA1}

\begin{figure*}[]
  \centering
  \includegraphics[width=\textwidth, angle=0]{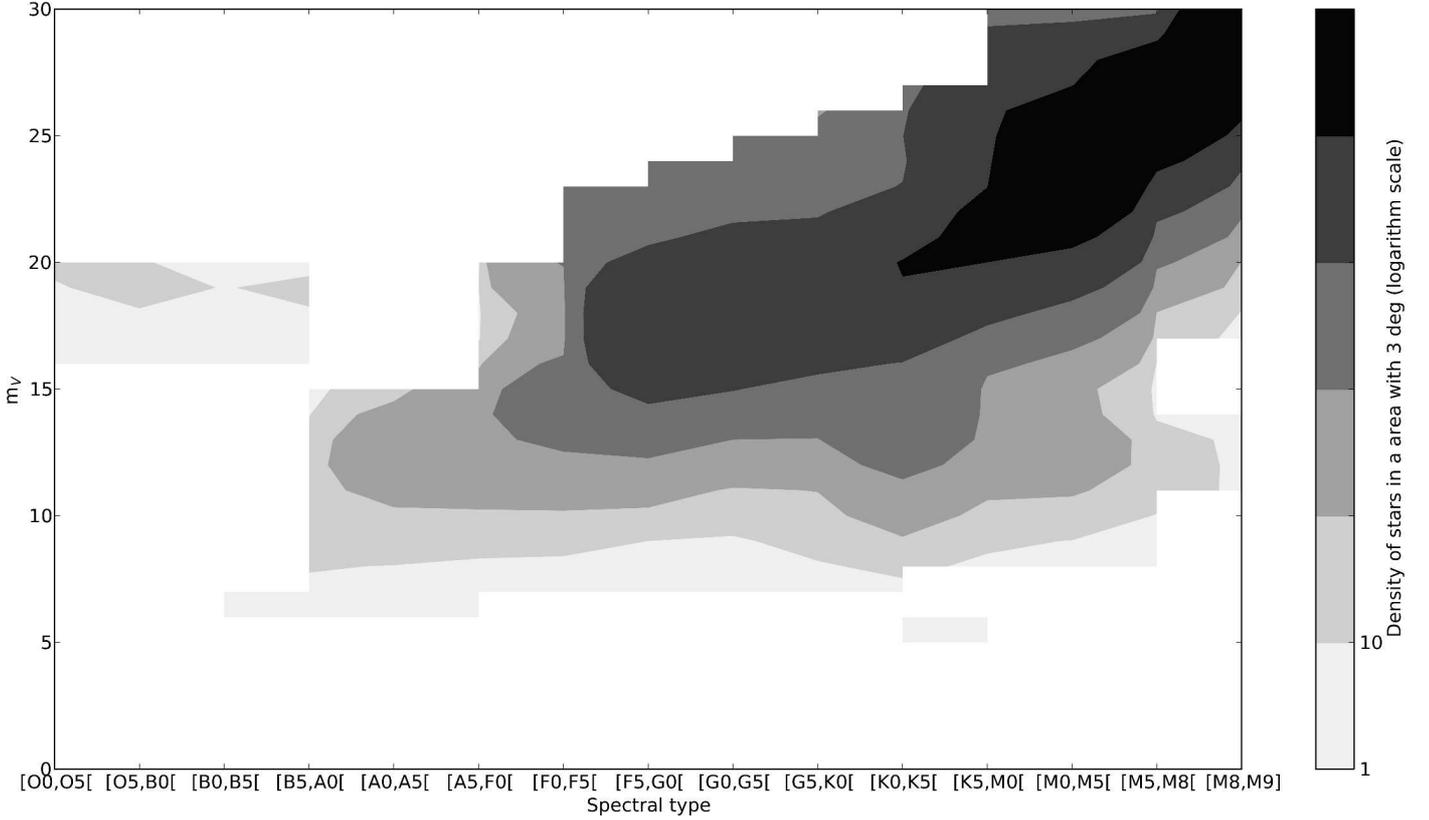}
  \caption{Number of stars in each bin of spectral type and magnitude (radius of $3^{\circ}$).}
  \label{tbl:3copper}
\end{figure*}

Even if the target star is a single star, an alignment with a background/foreground object may occur.
 To study the probability of such an event we used the Besan\c{c}on model
 \citep{Robin_et_al_2003}. 
This allows one to compute the probable stellar content on a given direction of the Galaxy, 
permitting one to infer the most probable contaminants. 
To do so, we simulated an observation 
in the direction of the  \textit{HD 85512} coordinates, $(l,b)=(271.6759,8.1599)$, which is positioned at the border 
of the Galactic disk. We also considered a distance interval of 50 kpc (approximately the diameter of the Milky Way), 
magnitudes between 0 and 30, and a solid angle for a radius of $3^{\circ}$. 
The objective was to have a large number of stars for each bin of magnitude and spectral type and then to divide 
it by the ratio between the area considered and that for the HARPS fiber (radius of 0.5 arcsec),
 thus obtaining the density of stars or the probability to have a fortuitous alignment within the HARPS fiber.
In Fig. \ref{tbl:3copper} we present a contour plot of the total number of
 stars obtained in the simulation for each bin
 of magnitude and spectral type. For more details on these values and the values for the star density when
 normalized for a HARPS size fiber, see  Appendix \ref{tables}. The most probable contaminants for the accidental alignments are  shown 
in Table \ref{tbl:moreprob}.

\begin {table}[]
 \caption{ Most probable contaminants for the fortuitous alignments.}
 \begin{center}
 \begin{tabular}{c c c}
   \hline
 \textbf{Spectral type} & \textbf{m$_V$} & \textbf{P [\%]}\\ \hline
 $[F0, F5[$ & $[13,16]\wedge[20,23]$&0.08\\
 $[F5, G0[$ & $[13,24]$ & 1.25\\
 $[G0, G5[$ & $[13,25]$ & 1.74\\
 $[G5, K0[$ & $[13,25]$ & 1.32\\
 $[K0, K5[$ & $[12,26]$ & 2.5\\
 $[K5, M0[$ & $[16,30]$ & 4.48\\
 $[M0, M5[$ & $[17,30]$ & 16.99\\
 $[M5, M8[$ & $[20,30]$ & 12.12 \\
 $[M8, M9]$ & $[21,30]$ & 13.17\\ \hline
 \end{tabular}
 \end{center}
 \label{tbl:moreprob}
 \end{table}

\section{Data}\label{sec:data}

We contemplated target stars from spectral type G to late M, with {visual} magnitudes, $m_v$, between 5 and 15. 
As described in the previous section (Sect. \ref {AA1}), the most probable contaminant stars are of spectral types FGKM. To represent
both target and contaminant spectral types, we chose spectra from nine stars, which are shown in Table  \ref{tbl:sn50}.

We used spectra with an S/N at  the center of the spectral order {number} 50 (which corresponds to a wavelength of 437.28 nm),
 varying between 15 for an M star, and 450 for a G star. 
The   S/N  values of these spectra can be found in Table \ref{tbl:sn50}.
The stars were chosen because they had the highest S/N in their spectral type range, as observed with HARPS.

 \begin {table}[]
 \caption{Stars used in our simulations, their spectral type, visual magnitude ($m_v$)  and  S/N at
 the center of the spectral order 50 (SN50).}
 \begin{center}
 \begin{tabular}{c c c c}
  \hline
\textbf{Object}& \textbf{Spec.Type}& \boldmath{$m_v$}$^{(1)}$ & \textbf{SN50} \\ \hline
HD20852     & F2 & 6.92 &225; 98 \\
HD103774    & F5 & 7.12 &102; 146\\
Sun         & G2 & {3.65} $^{(2)}$  &225; 260\\
Tau Ceti    & G8 & {3.50} &300; 450\\
HD69830     & K0 & {5.95} &253; 380\\
Alpha Cen B & K1 & {1.33} & 385; 399; 302\\
HD85512     & K5 & {7.65} & 156; 86\\
Gl436       & M1 & {10.68} & 15; 31; 46\\
Gl581       & M3 & {10.57} & 37; 22\\\hline
 \end{tabular}
 \end{center}
 \label{tbl:sn50}
$^{(1)}$ from Simbad (http:simbad.u-strasbg.fr/simbad/sim-fid)\\
$^{(2)}$ $m_v$ of the moon within the HARPS 0.5'' radius fiber.
 \end{table}

\section{Method}\label{sec:Met}

The goal of this work is to study possible stellar contamination of a target star. To do so,
we co-added target and contaminant spectra.
We not only 
considered contaminant stars of spectral types F2, F5, G2, G8, K0/K1, K5, M1, and M3, but also different ratios of 
magnitude between target star and contaminant as well as differences in RV. To sum two spectra 
with these particular conditions is not trivial. We need to  control  the magnitudes and RVs. 

 We set the magnitude V of the target ($m_t$) star according to the 
data available in Simbad \footnote{http://simbad.u-strasbg.fr/simbad/sim-fid} and then calculated the
 magnitude of the contaminant ($m_c$) with respect to the target star using the apparent magnitude - flux relation.
We also considered the relative RV between the two spectra as a tunable knob. However, when processing our spectra using the HARPS
pipeline the reduction takes into account the header (and properties) of the parent spectra. 
To overcome this situation we calculated the contaminant RV in the target reference frame and  shifted
it afterwards. 

We then co-added the target spectrum  with the modified contaminant spectra
and processed the result with the HARPS RV pipeline. 

More details of the method are presented in Appendix \ref{details}.
 
\subsection{The ``real binaries'' star strategy}
To study the impact of a companion on the RV calculation of a binary primary star, we reproduced the cases discussed in  the work of 
\citet{Duquennoy_Mayor_1991}. We considered a primary G2 star
with six different secondary/contaminant stars: G2, G8, K1, K5, M1, and M3.

\begin{figure}[]
  \centering
  \includegraphics[width=\columnwidth]{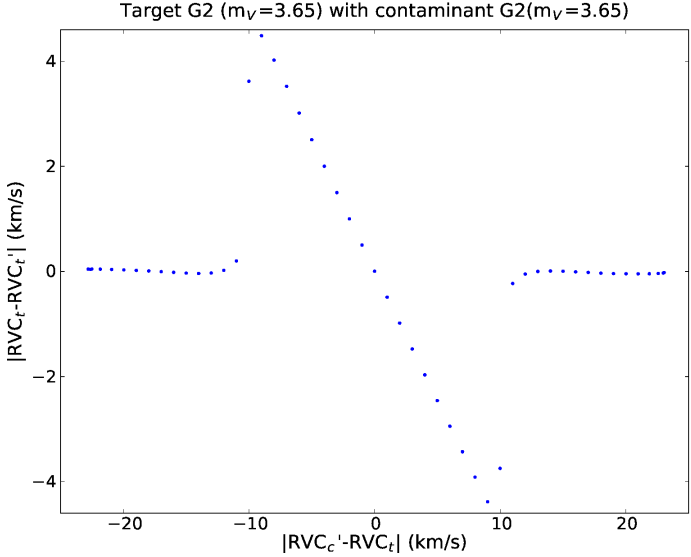}
  \caption{Shift of a G2 target RV caused by a contaminant of the same spectral type and magnitude, $m_v$, in function of the 
 difference between individual radial velocities.}
  \label{fig:G2_5.15_G2_5.15total}
\end{figure}

 The first case had a G2 as contaminant with positive and negative values of 
$\Delta RV = RVC_c'-RVC_t$, where $RVC_t$ is the target  barycentric radial velocity (drift corrected) without contamination,
 and $RVC_c'$ is the contaminant barycentric radial velocity (drift corrected) in the target reference frame. 
  In Fig. \ref{fig:G2_5.15_G2_5.15total} we show the impact, $RVC_t-RVC_t'$, of the stellar companion on the target RV, 
where $RVC_t'$ is the target barycentric radial velocity (drift corrected) with contamination.

For each target-contaminant pair we
considered a range of $|\Delta RV|$ between 0 and 30[Km/s] (in steps of 1 [Km/s]), except for 
 a secondary star of spectral type K5 in which the maximum  $|\Delta RV|$ was of 24 [Km/s] (see Appendix \ref{details} for details).

\subsection{Fortuitous alignment strategy}
The fortuitous alignment case is more generic. Here we considered six spectral types of primary stars (G2, G8, K0, K5, M1, and M3) in 
combination with the eight spectral types as contaminants -- the same six used as target stars plus an F2 (HD20852) and an F5 (HD103774) star. 
For each spectral type of primary stars we 
considered contaminants with magnitudes $mv$ between 5 and 15 (in steps of one). The contaminants considered were those described 
 in Table \ref{tbl:moreprob}.  

Although the Besan\c{c}on model includes oxygen and carbon AGB (asymptotic giant branch) stars and  
 white dwarfs in the search, we only considered main-sequence (MS) 
stars. Very metal-poor stars were also excluded from our study.
 For  an M1 target star a K0 star (HD69680) was used instead of the K1 mentioned above. 
This was because the barycentric Earth radial velocity (BERV) value  form the K1 star did not allow the desired shift in RV.
 For each target-contaminant pair we considered the 
differences between its RV, $|\Delta RV|$, under 20 Km/s, which is half the width of the CCF window. 
Although giant stars will often be contaminant stars, we assumed that spectra from MS stars are similar to those of giants and 
used only spectra of MS stars in our simulations.
\section{Results}\label{sec:Resul}

\subsection{Real binaries}\label{sec:BinR}

\begin{figure*}[]
  \centering
  \includegraphics[width=\textwidth]{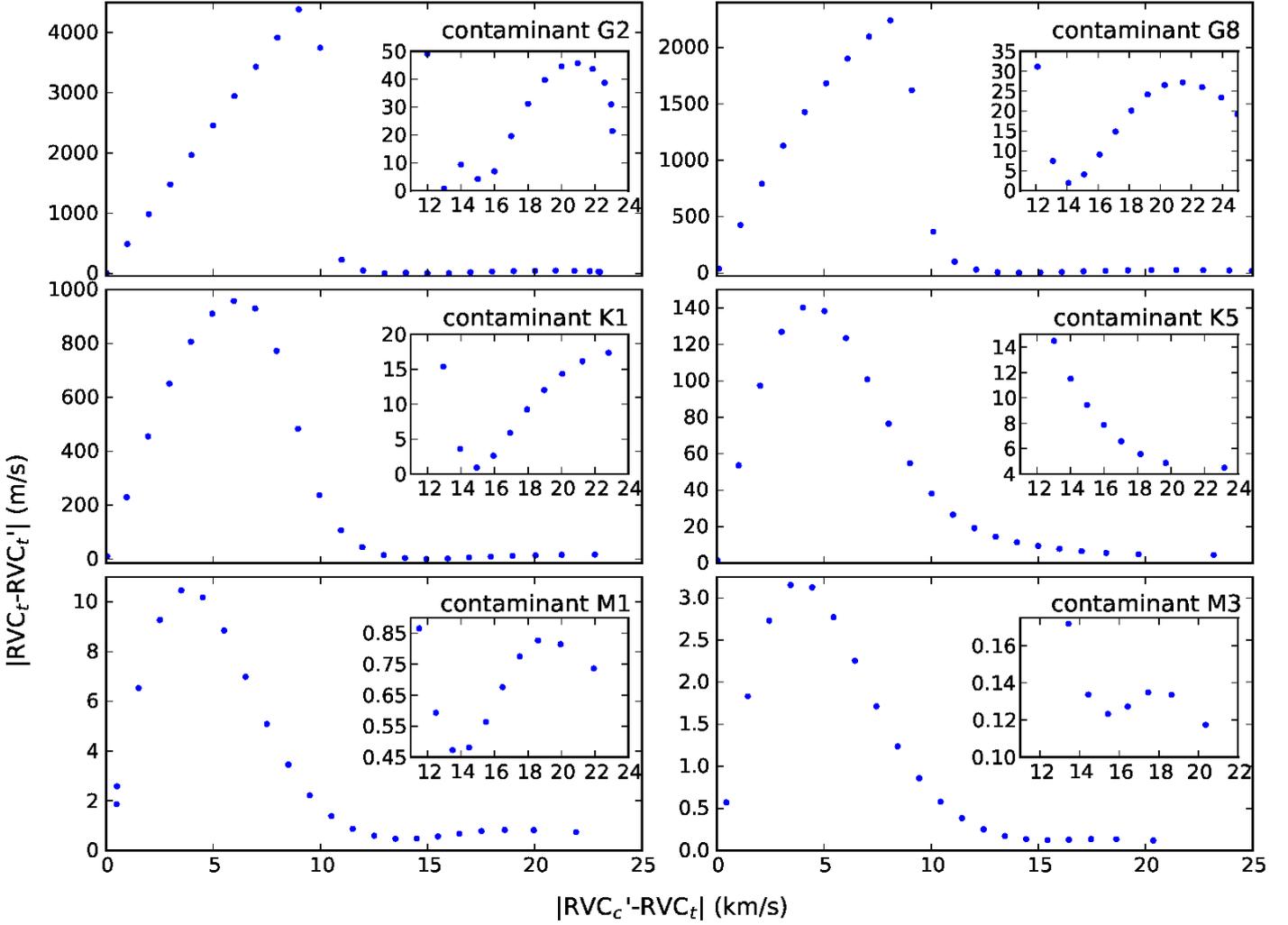}
  \caption{Impact on the RV of a G2 target star caused by a companion of
 spectral type G2, G8, K1, K5, M1, and M3 (left to right, top to bottom). 
For each case a close-up with a bump for the impact for $10<\Delta RV  <24$ $km/s$ is shown. }
  \label{fig:BinarioReal}
\end{figure*}

The results for the contamination by the secondary star of a binary with a primary star of spectral type G2 
as a function of $\Delta RV$ can be found in Fig. \ref{fig:BinarioReal}.
Obviously, the impact on the RV calculation is a function of the difference between target and contaminant RV ($\Delta RV$).
One can interpret a CCF as an ``average'' line of a spectrum, that is fitted by a Gaussian curve  to measure the center of the CCF. 
 In the presence of a second star, the CCF is composed of two curves. For  $\Delta RV=0$ the two curves  fully overlap, and so the impact 
on the Gaussian fit and on the RV calculation is minimum. As $\Delta RV$ increases, the impact on the Gaussian fit increases until
the point
 when the two CCF curves  overlap, with the contaminant RV, $|RVC'_c|$, close to 0.6 times the FWHM of the target CCF.
When the CCF curves become distinct, the impact on the measured RV of the primary decreases.
 For $|\Delta RV|>20\, km/s$ the peak of the contaminant CCF curve exits 
the CCF window that is usually computed with a value of 20 $km/s$, and for values $|\Delta RV|\geq25 \, km/s$ the contaminant 
CCF curve completely exits the target CCF 
window. Nevertheless, the contamination is not null. There are bumps in the  wings that slightly contaminate the target
spectra.   

\begin{figure}[]
  \centering
  \includegraphics[width=\columnwidth]{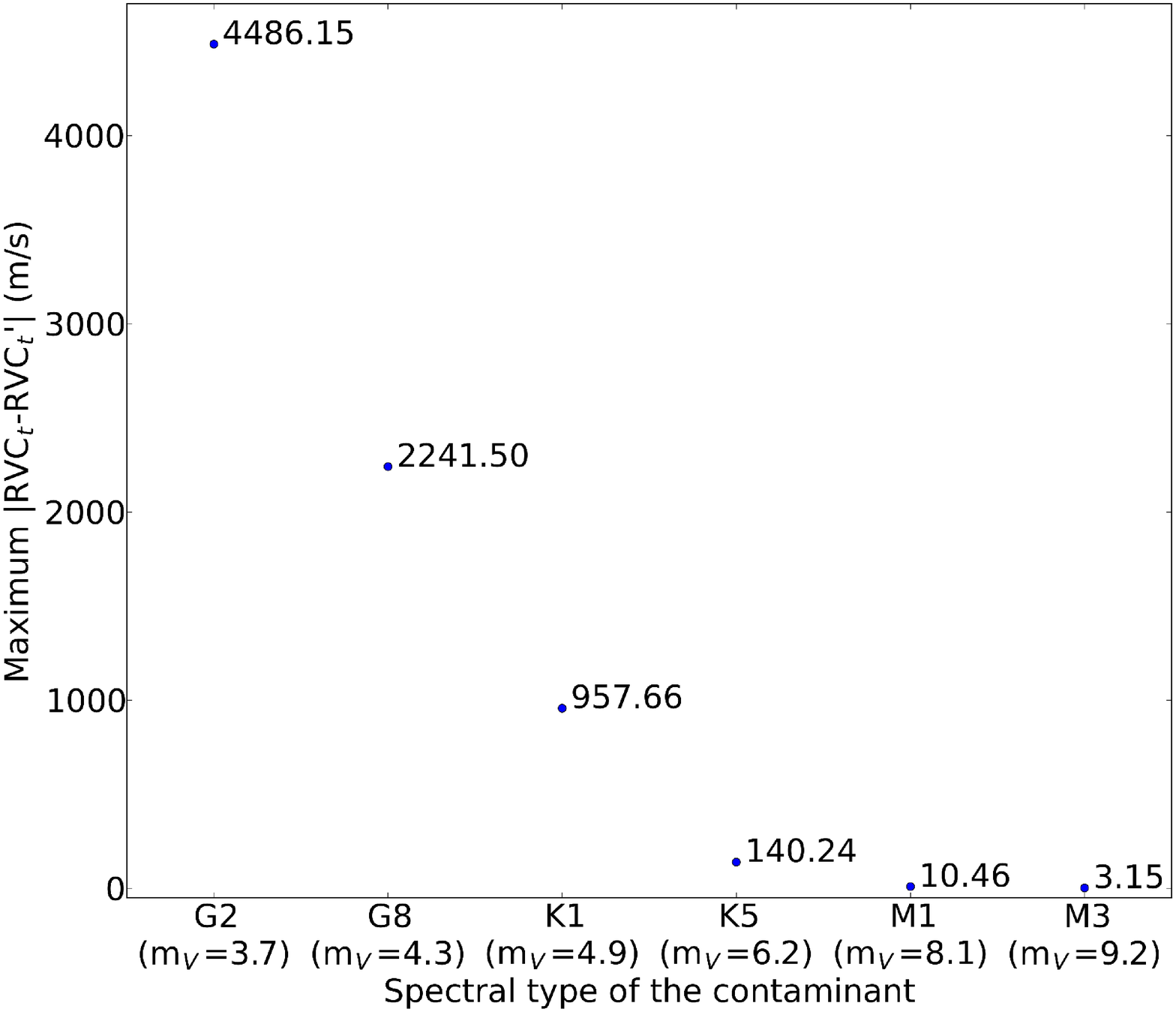}
  \caption{Maximum contamination of a G2 target star caused  by each companion spectral type in m/s.}
  \label{fig:Maximos_binarios}
\end{figure}

 Figure \ref{fig:Maximos_binarios}  shows the maximum contamination for each type of contaminant.
As we can see, as the secondary stars become fainter and of later spectral type, the maximum contamination
decreases. 

\subsection{Contamination by the sky}
The sky brightness is not zero and its light results from the reflection and scattering of Sun spectra. 
The sky brightness has a magnitude per arcsec$^2$  in V band 
of 21.8 on a new-moon night, and of 20.0 on a full-moon night.
To study the impact of this contamination we assumed that sky spectrum is the solar one. 
This is equivalent to say 
that for a fiber of 0.5'' radius, we have a systematic contamination of a G2 star with m$_V\sim20$ on a full-moon night 
to a limit of a G2 star with  m$_V\sim22$  (new moon).  

\begin{figure*}[]
\begin{center}
\mbox{
\subfigure[]{\resizebox{0.418\textwidth}{!}{\includegraphics{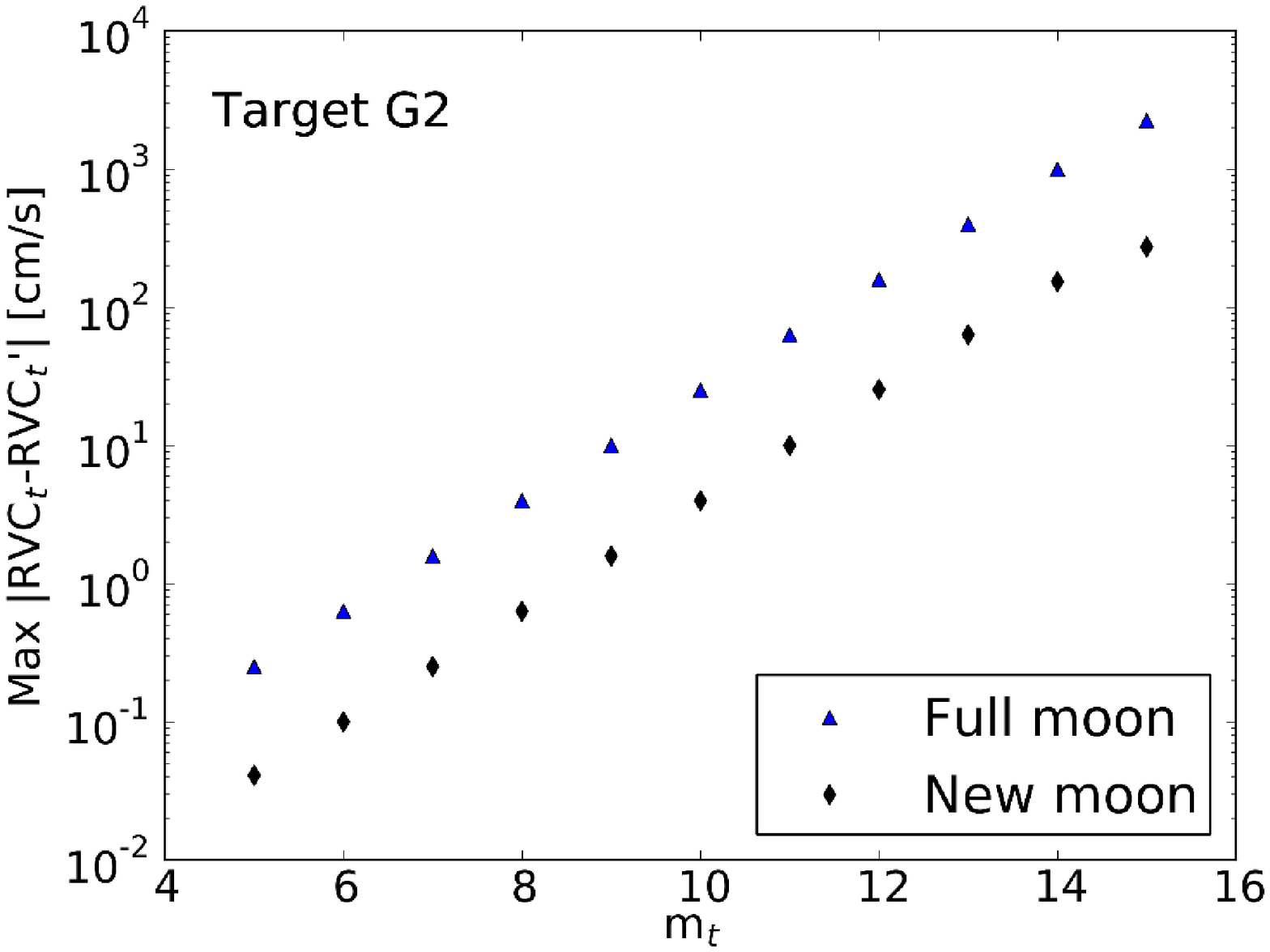}\label{figLuaG2_max}}}\hspace{0.00cm}}
\mbox{\hspace{0.0cm}
\subfigure[]{\resizebox{0.418\textwidth}{!}{\includegraphics{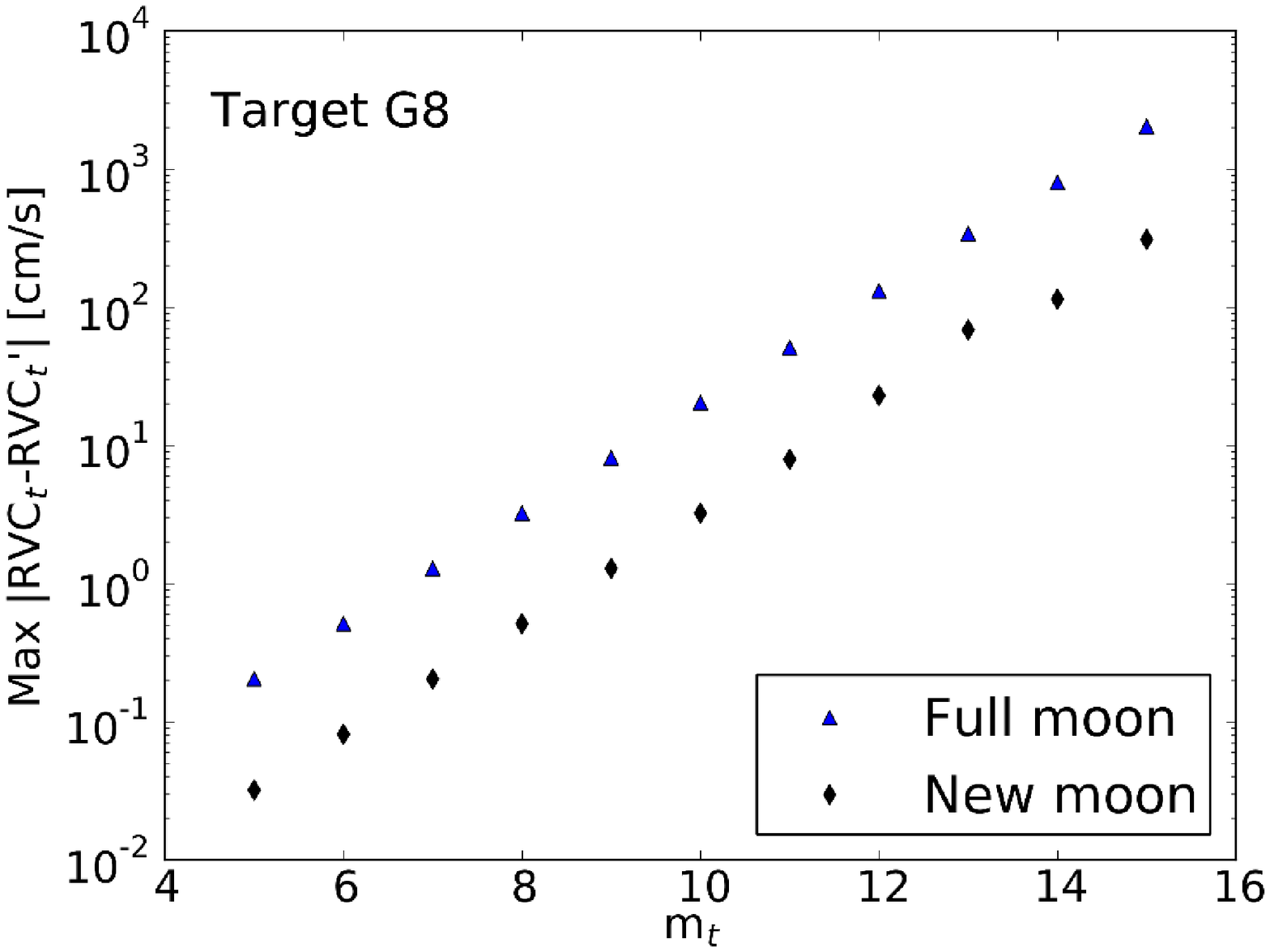}\label{figLuaG8_max}}}\hspace{0.00cm}}
\mbox{\hspace{0.0cm}
\subfigure[]{\resizebox{0.418\textwidth}{!}{\includegraphics{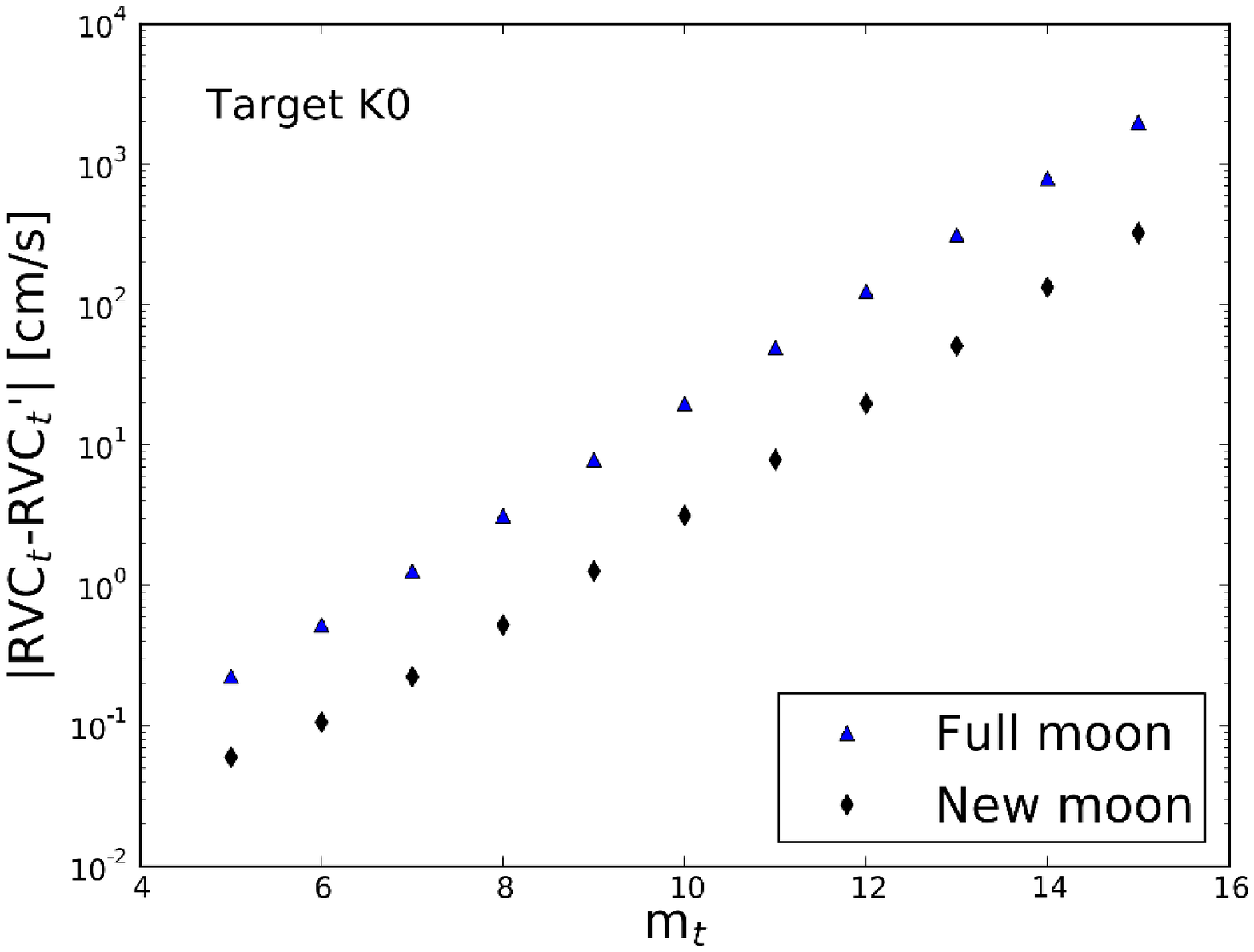}\label{figLuaK0_max}}}\hspace{0.00cm}}
\mbox{\hspace{0.0cm}
\subfigure[]{\resizebox{0.418\textwidth}{!}{\includegraphics{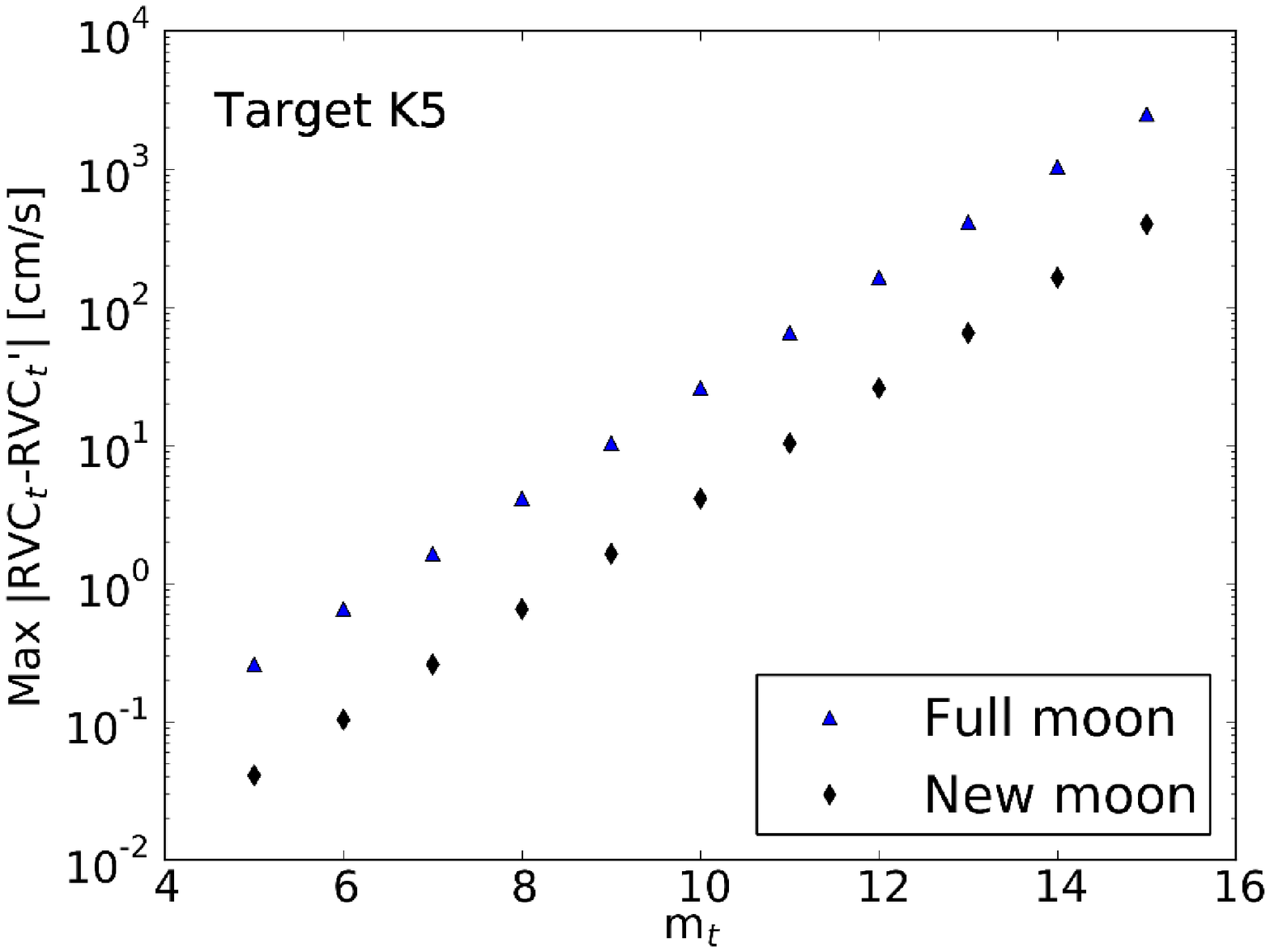}\label{figLuaK5_max}}}\hspace{0.00cm}}
\mbox{\hspace{0.0cm}
\subfigure[]{\resizebox{0.418\textwidth}{!}{\includegraphics{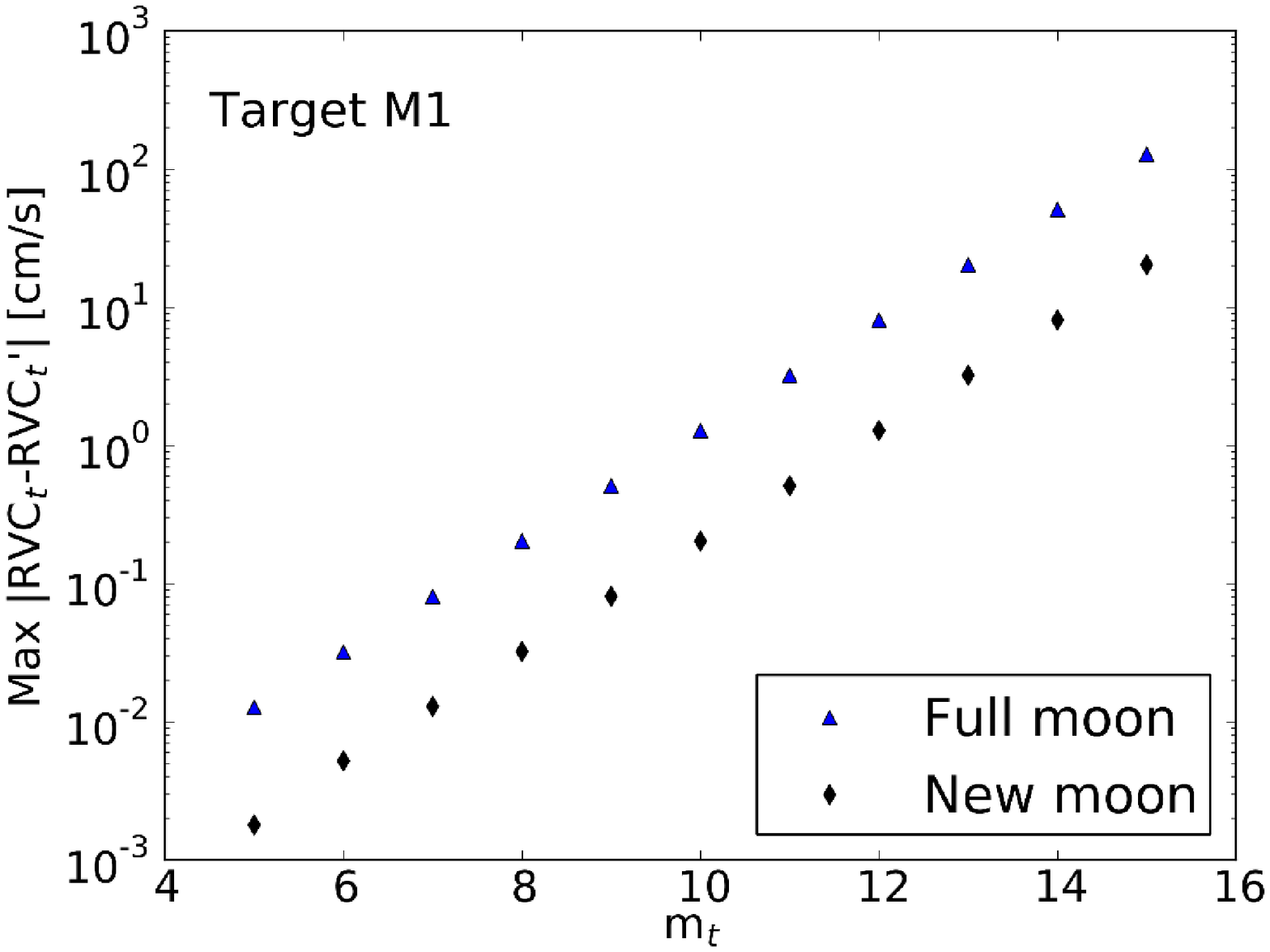}\label{figLuaM1_max}}}\hspace{0.00cm}}
\mbox{\hspace{0.0cm}
\subfigure[]{\resizebox{0.418\textwidth}{!}{\includegraphics{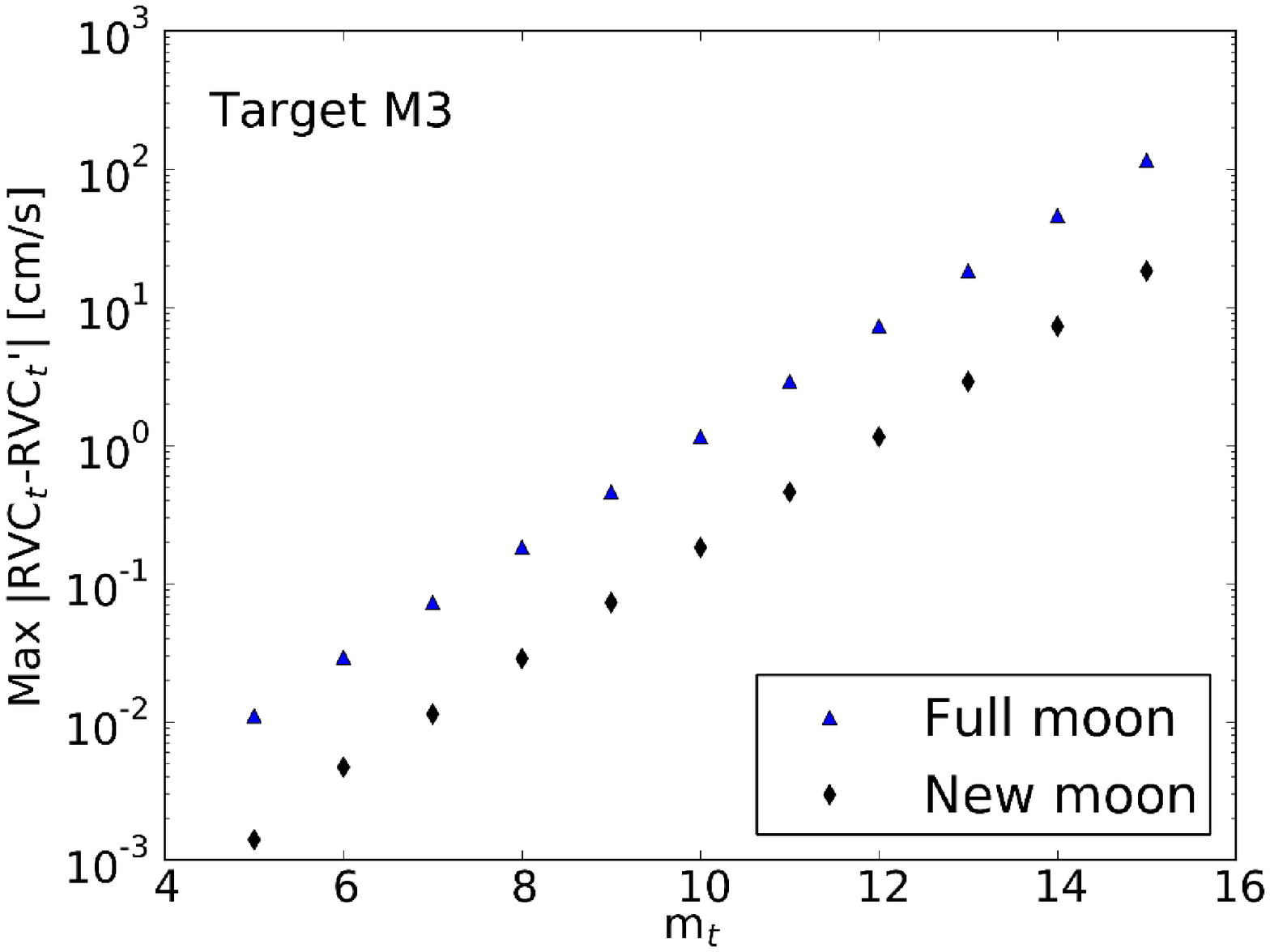}\label{figLuaM3_max}}}}
\caption{\label{figLua_max}Maximum contamination of the  sky brightness on a \textbf{a)} G2, \textbf{b) }G8,
 \textbf{c) }K0, \textbf{d)} K5, \textbf{e)} M1, and \textbf{f)} M3 target star spectra. The blue triangles show the impact
 on the RV caused by the sky brightness on a full-moon night, while bçlack diamonds show the impact on a new-moon night.}
\end{center}
\end{figure*}

The results of the maximum contamination by the sky brightness study are shown in  Fig. \ref{figLua_max}.
We can see that when observing a G2, G8, K0, or a K5 target star fainter that m$_V=11$, or  M1, M3 fainter that m$_V=14$, 
the errors induced on the observed RV can be larger than 10 cm/s, 
which can be a problem for future instruments such as ESPRESSO.
For a night of observation with a full moon, induced errors become larger than 10 cm/s for 
G- and K target stars fainter that m$_V=9$ and for an M star fainter than m$_V=12$.

\subsection{Fortuitous alignment} \label{sec:FortR}

\begin{figure}[]
  \centering
  \includegraphics[width=0.45\textwidth]{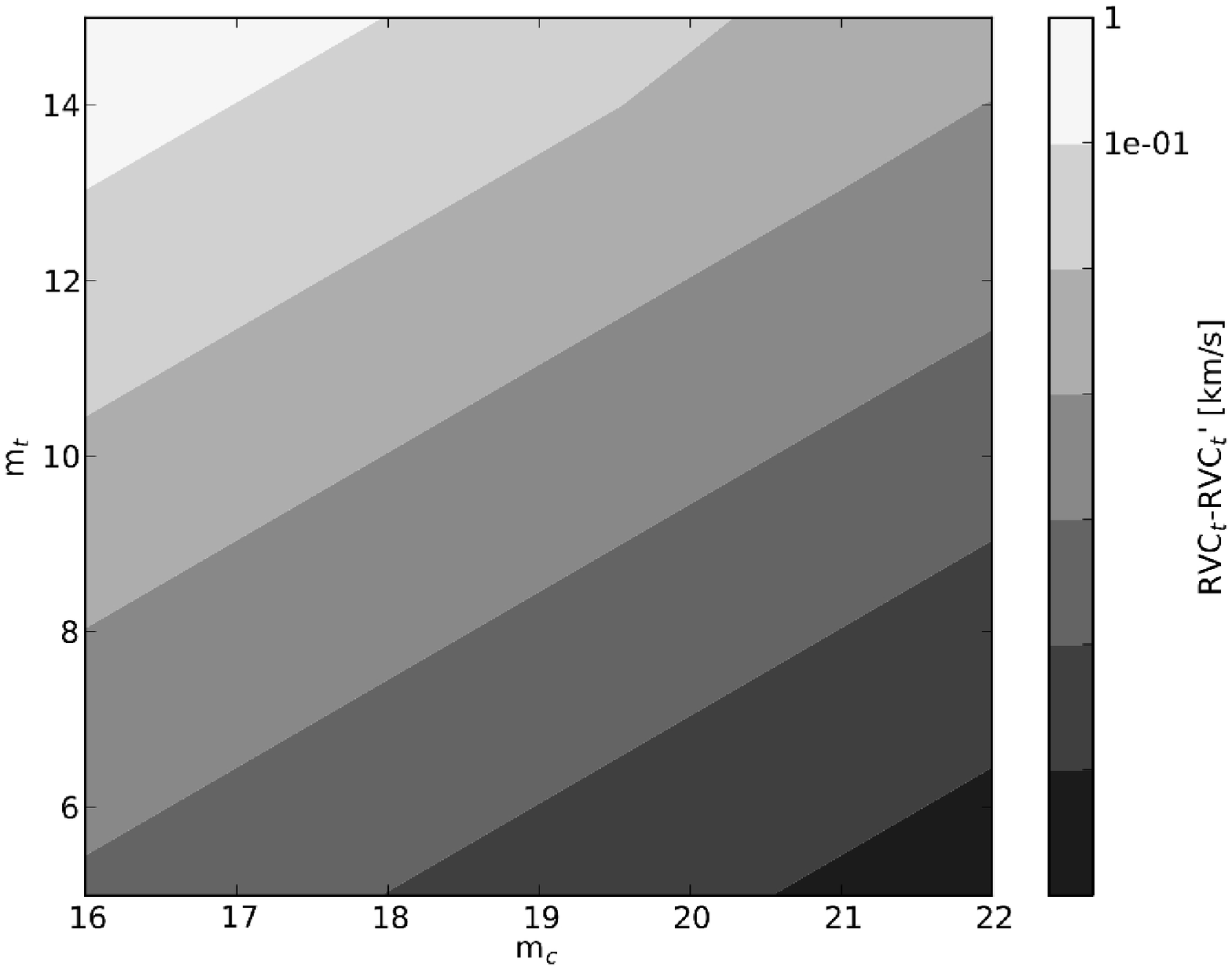}               
  \caption{{Contour plot of the maximum induced RV shift in the target 
$|RVC_t - RVC_t'|$ depending on the target (m$_t$) and contaminant (m$_c$) magnitudes.}}
  \label{fig:contour}
\end{figure}

\begin{figure*}[]
\begin{center}
\mbox{
\subfigure[]{\resizebox{0.45\textwidth}{!}{\includegraphics{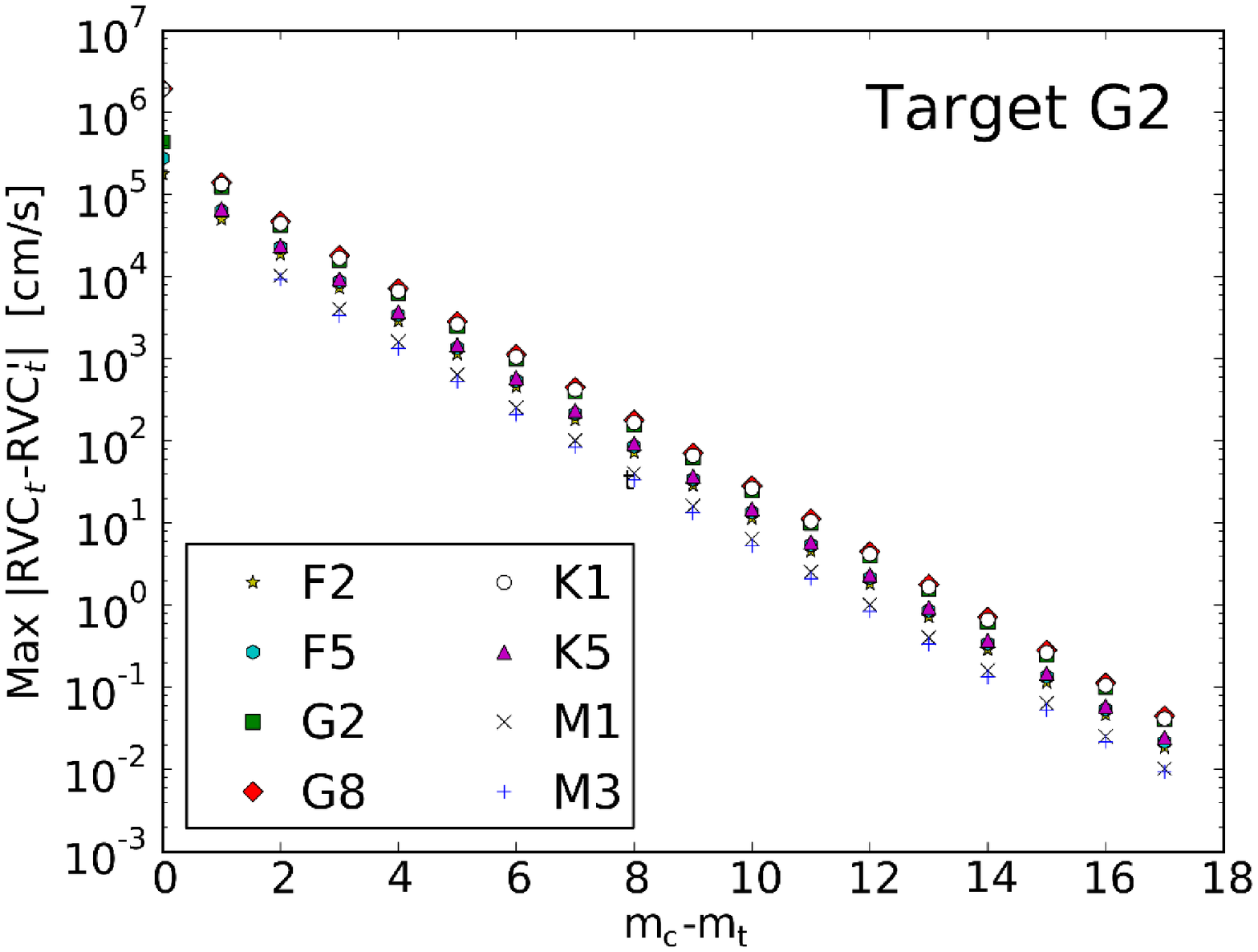}\label{fig:G2comum}}}\hspace{0.00cm}}
\mbox{\hspace{0.0cm}
\subfigure[]{\resizebox{0.45\textwidth}{!}{\includegraphics{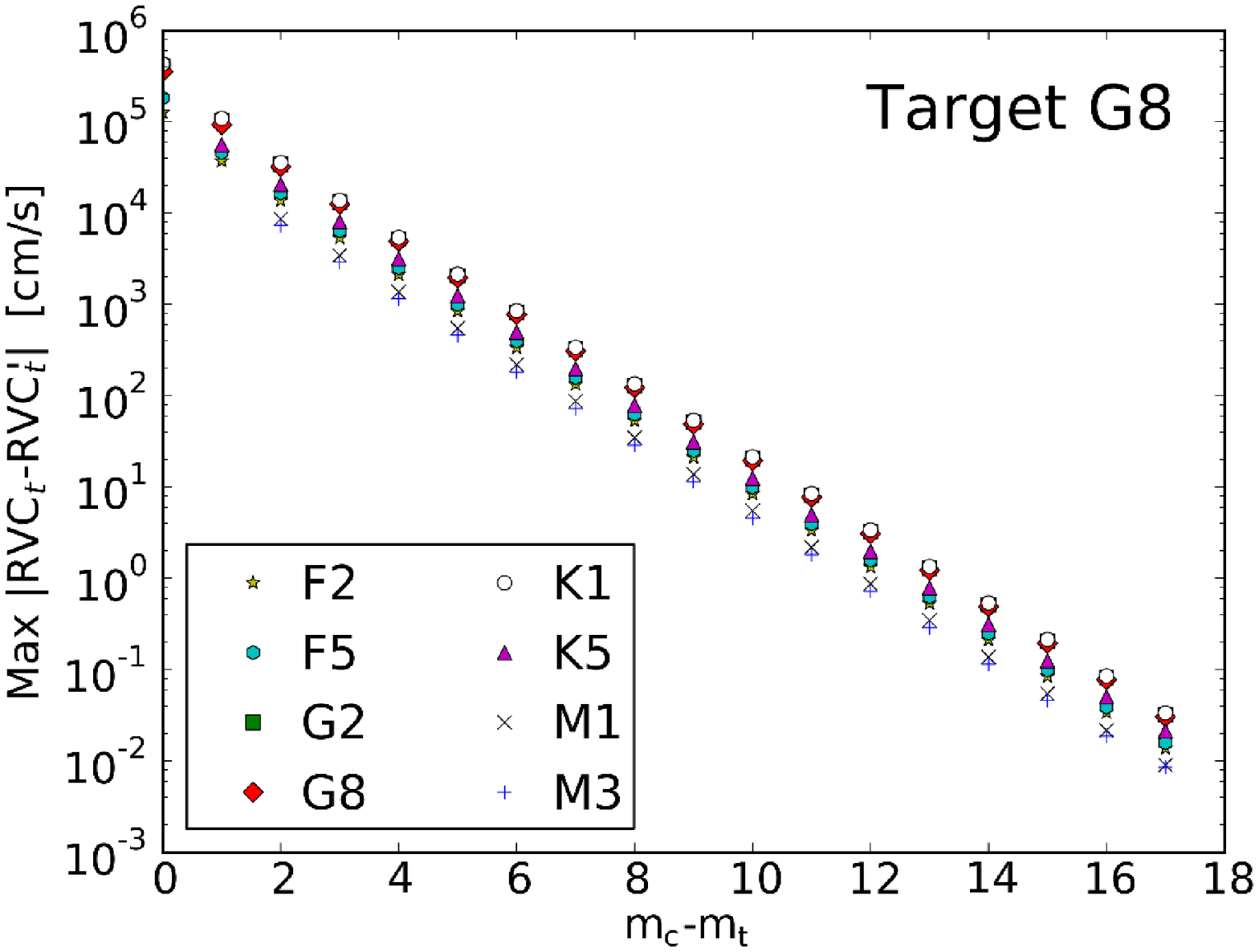}\label{fig:G8comum}}}\hspace{0.00cm}}
\mbox{\hspace{0.0cm}
\subfigure[]{\resizebox{0.45\textwidth}{!}{\includegraphics{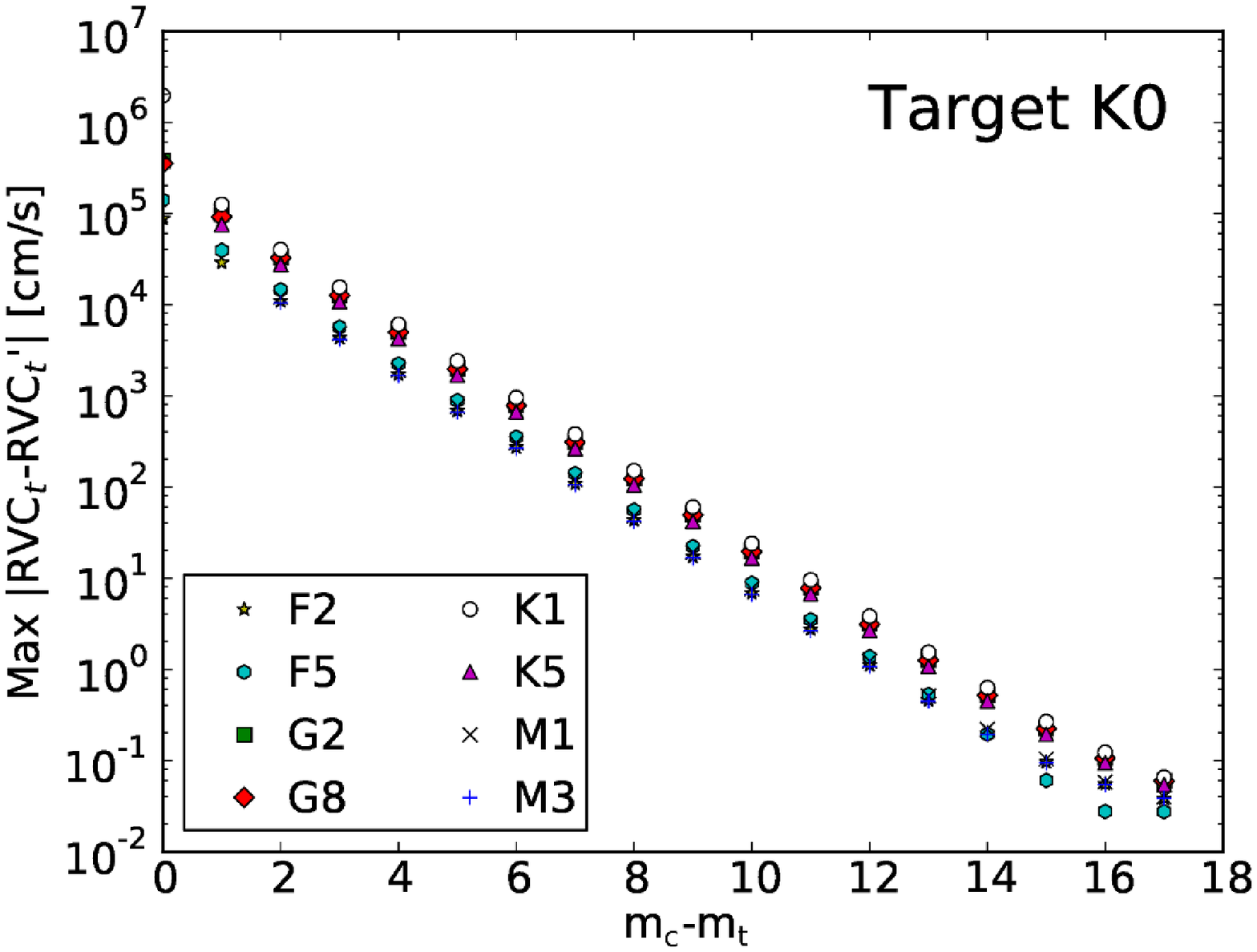}\label{fig:K0comum}}}\hspace{0.00cm}}
\mbox{\hspace{0.0cm}
\subfigure[]{\resizebox{0.45\textwidth}{!}{\includegraphics{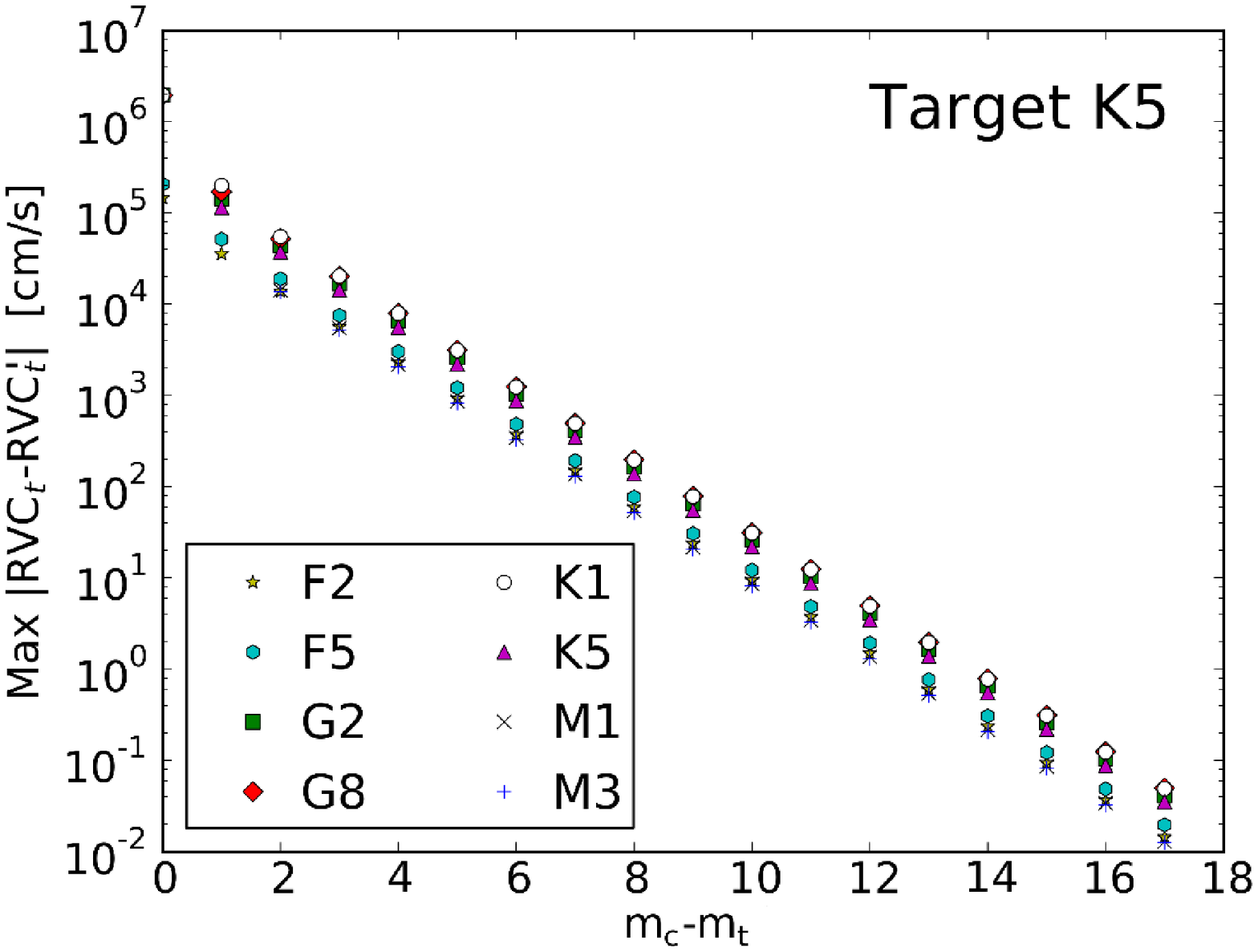}\label{fig:K5comum}}}\hspace{0.00cm}}
\mbox{\hspace{0.0cm}
\subfigure[]{\resizebox{0.45\textwidth}{!}{\includegraphics{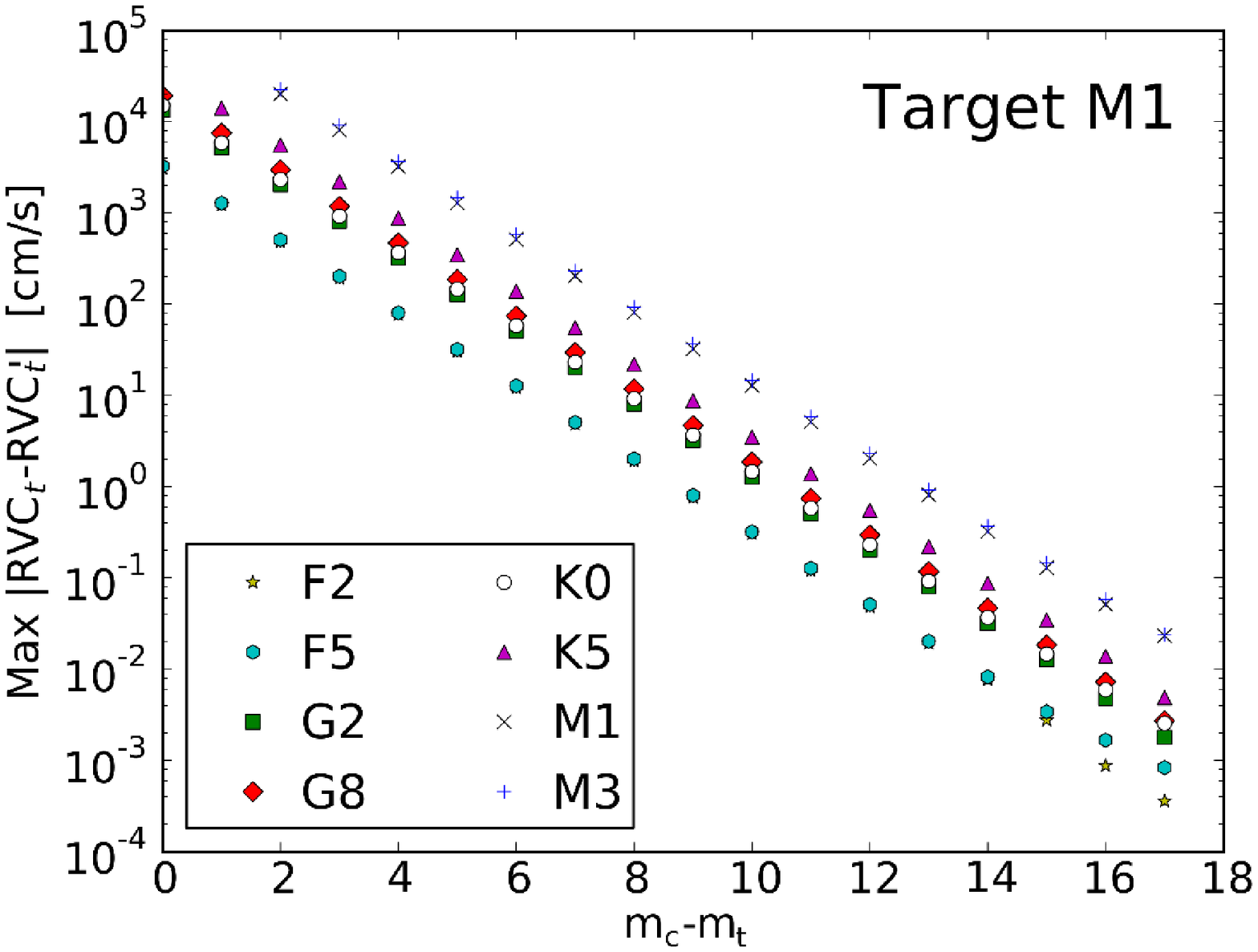}\label{fig:M1comum}}}\hspace{0.00cm}}
\mbox{\hspace{0.0cm}
\subfigure[]{\resizebox{0.45\textwidth}{!}{\includegraphics{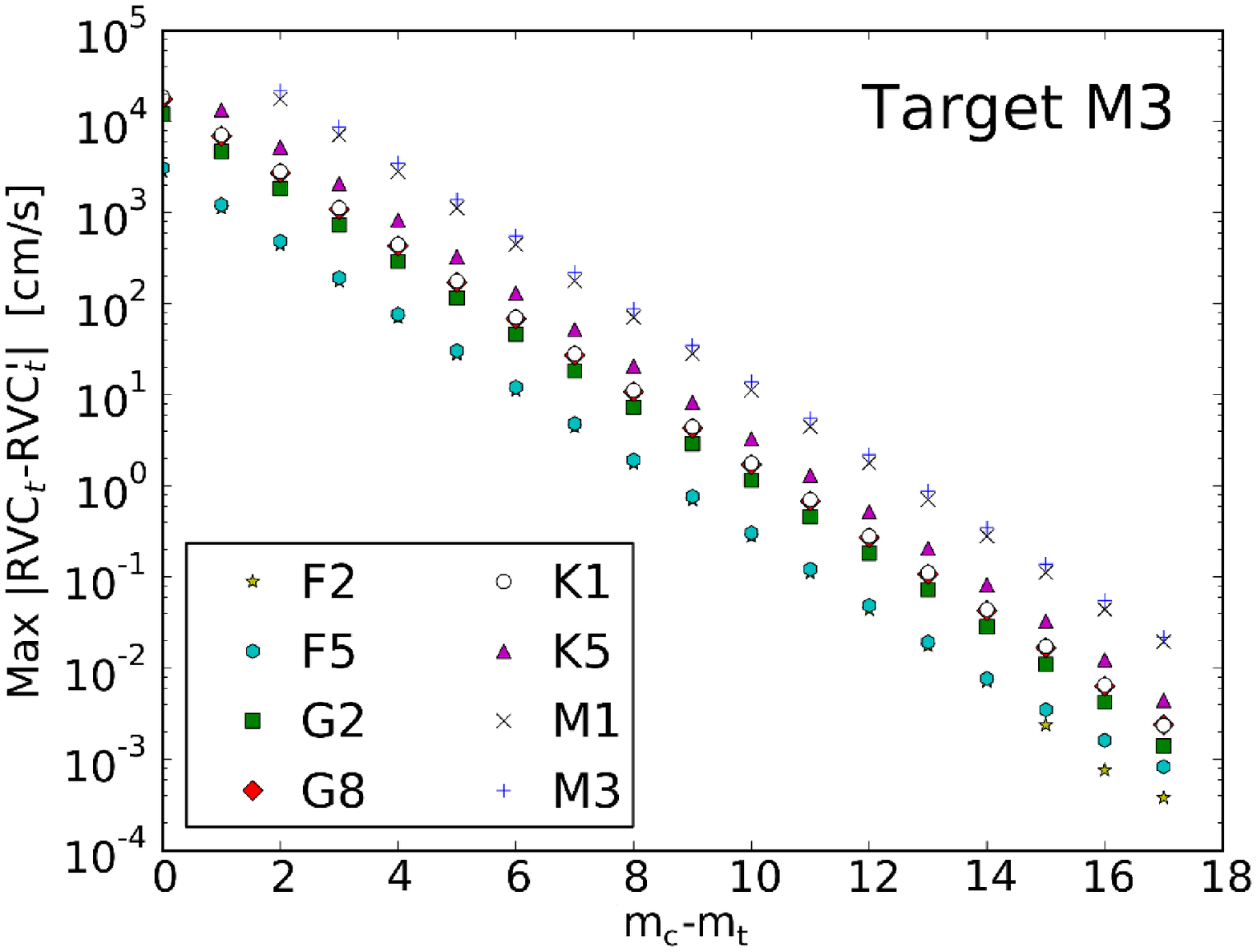}\label{fig:M3comum}}}}
\caption{\label{fig:comum}Maximum impact of $|\Delta m|$ on the $|RVC_t|$ for a target star of spectral type \textbf{a)} G2, \textbf{b) }G8,
 \textbf{c) }K0, \textbf{d)} K5, \textbf{e)} M1, and 
 \textbf{f) }M3, and  contaminants of spectral types F2 (star), F5 (hexagon), G2 (squares), G8 (diamonds), K1 (circles), K5 (triangles), M1 ( x crosses), and 
 M3 (+ crosses).}
\end{center}
\end{figure*}

In this section we present a more exhaustive study of all  possible sources of stellar contamination.
Beyond  the moon as a contaminator, there is a vast range of possible contaminators, 
as was discussed in Sect. \ref{AA1}. Our study is simplified because our analysis holds for similar 
spectra with different magnitude for target and contaminant star, but with same difference between them 
($\Delta m=m_c-m_t$), as we can see in Fig. \ref{fig:contour}. This way, we only carried out our study on the impact for a 
target star in which m$_v$ differs $|\Delta m|$ from the contaminant star. The result of our study for contamination as a function of spectral type
 is shown in Fig. \ref{fig:comum}. 

\begin{figure}[]
  \centering
  \includegraphics[width=\columnwidth, angle=0]{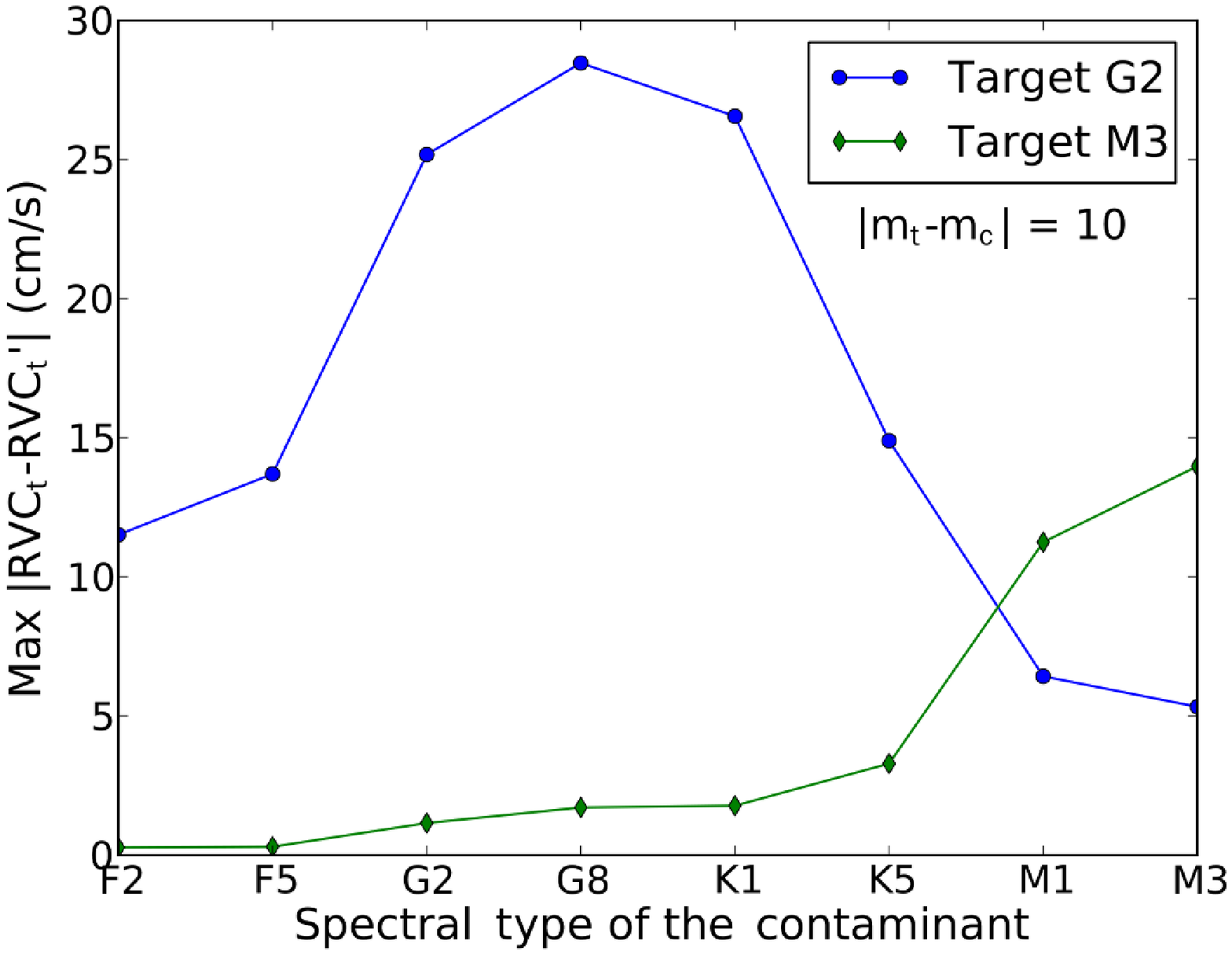}
  \caption{Maximum contamination $|RVC_t -RVC_t'|$ as a function of the contaminant spectral type for 
a G2 (blue dots) and M3 (green diamonds) target star with $\Delta m =10$.}
  \label{fig:RVCSpecType}
\end{figure}

We first note that there is a clearly defined 
relation between $\Delta m$ and the maximum of the contamination, except for some cases where the difference in magnitude between 
target and contaminant is null. 
A misidentification of the companion correlation peak with that of the primary may happen due to a (slightly) larger contrast.
 If we take only $\Delta m > 1$,  we can describe the relation between 
 $\Delta m$ and the maximum of the contamination  as
\begin{equation}
\mathrm{Max} |RVC_t - RVC_t'| = 10^{a\cdot\Delta m+b},  
\end{equation}
 where $a$ and $b$ are the coefficients presented in Table \ref{tbl:fit}. In this table we also present the value for the 
root mean square (rms) for each fit. By comparing the panels of  Fig. \ref{fig:comum} we can also see that the contamination
 depends not only on $\Delta m$, but also on the  target-contaminant spectral type combination. This becomes more evident in
Fig. \ref{fig:RVCSpecType}. This figure shows the maximum contamination as a function of the contaminant spectral type for 
a G2 and M3 target star with $\Delta m =10$. We find two reasons for this behavior. 
First, less contamination occurs with increasingly different spectra.  Also, the contamination will 
 depend on the depth of spectral line. Thus, for similar
spectra, cooler stars will cause a higher contamination, because their spectral lines depth is higher. 

\begin {table}[]
  \footnotesize
  \centering
  \caption{Coefficients $a$ an $b$ for each fitting of the relation between 
$\Delta m$ and the maximum impact on the RV calculations: $\mathrm{Max} |RVC_t - RVC_t'| = 10^{a\cdot\Delta m+b}$, and rms resulting from 
each combination of target spctral type (TST) and contaminant spectral type (CST).} \label{tbl:fit}
 \begin{tabular}{c c c c c c }
\toprule 
 \textbf{TST } &  \textbf{CST } & \textbf{a } & \textbf{b } & \textbf{rms} \\
 &   &             &             & \textbf{[cm/s] }\\ \midrule
G2&     F2      &   -0.4008    & 5.0733         &   815.0813 \\ 
  &     F5      &   -0.4019    & 5.1601         &  1644.9045 \\ 
  &     G2      &   -0.4022    & 5.4288         &  4491.1375 \\
  &     G8      &   -0.4023    & 5.4829         &  4956.4768 \\
  &     K1      &   -0.4026    & 5.4556         &  5136.0605 \\
  &     K5      &   -0.4009    & 5.1859         &  1500.0546 \\
  &     M1      &   -0.4001    & 4.8096         &    23.3793 \\
  &     M3      &   -0.3996    & 4.7301         &  197.1604  \\
G8&     F2      &   -0.4008     & 4.9379        &  681.0614  \\
  &     F5      &   -0.4019     & 5.0222        & 1075.0923  \\
  &     G2      &   -0.4027     & 5.3440        &  3968.9098 \\
  &     G8      &   -0.4021     & 5.3154        &  2697.4956 \\
  &     K1      &   -0.4027     & 5.3641        &  4215.8544 \\
  &     K5      &   -0.4007     & 5.11          &  1139.4446 \\
  &     M1      &   -0.3995     & 4.7357        &     3.0972 \\
  &     M3      &   -0.3984     & 4.651         &    51.5545 \\ 
K0&     F2      &   -0.3812     & 4.7320        &  1566.6309 \\
  &     F5      &   -0.4008     & 4.9521        &   871.3506 \\
  &     G2      &   -0.3933     & 5.2712        &  5307.563 \\
  &     G8      &   -0.3929     & 5.2613        &  4246.1257 \\
  &     K1      &   -0.3954     & 5.3704        &  7390.8788 \\
  &     K5      &   -0.3914     & 5.1827        &  3347.1766 \\
  &     M1      &   -0.3799     & 4.7540        &   566.5518 \\
  &     M3      &   -0.3777     & 4.6783        &   532.1239 \\
K5&     F2      &   -0.3988     & 4.9559        &   200.036 \\
  &     F5      &   -0.3999     & 5.0865        &   724.4446 \\
  &     G2      &   -0.4035     & 5.4592        &  7366.4185 \\
  &     G8      &   -0.4032     & 5.5354        &  8587.4134 \\
  &     K1      &   -0.4053     & 5.5580        & 14146.5377 \\
  &     K5      &   -0.4030     & 5.3793        &  4935.42 \\
  &     M1      &   -0.4011     & 4.9458        &    71.0313 \\
  &     M3      &   -0.4009     & 4.9257        &    98.6394 \\ 
M1&     F2      &   -0.4056     & 3.5177        &    18.1497 \\
  &     F5      &   -0.3932     & 3.4673        &    24.6295 \\
  &     G2      &   -0.4018     & 4.1212        &    10.1997 \\
  &     G8      &   -0.4011     & 4.2786        &    13.0058 \\
  &     K1      &   -0.3990     & 4.1604        &    24.4166 \\
  &     K5      &   -0.4014     & 4.5525        &    19.9346 \\
  &     M1      &   -0.3985     & 5.0993        &    33.3093 \\
  &     M3      &   -0.3994     & 5.1581        &    56.2493 \\ 
M3&     F2      &   -0.4052     & 3.4770        &    14.7829 \\
  &     F5      &   -0.3924     & 3.4421        &    25.7393 \\
  &     G2      &   -0.4035     & 4.0854        &    30.8372 \\
  &     G8      &   -0.4019     & 4.2469        &    23.5459 \\
  &     K1      &   -0.4024     & 4.2649        &    35.7392 \\
  &     K5      &   -0.4021     & 4.5317        &    32.2298 \\
  &     M1      &   -0.3991     & 5.0440        &    13.2551 \\
  &     M3      &   -0.3999     & 5.1436        &    63.9946 \\ 
\bottomrule
 \end{tabular}
 \end{table}

\subsection{Statistical analysis of the contamination}\label{sec:SA}
After calculating the impact caused by a contaminant star in the RV calculation, it is important to known how often this occurs.
Therefore we calculated the probability of measuring the target RV with a  given contamination.

In the previous two subsections (Sects. \ref{sec:BinR} and \ref{sec:FortR}) we presented a study for the impact  to have
a contaminant star within the fiber on the RV calculations, assuming that the fiber was well centered on target and contaminant. 
Although this is a relevant study, the most probable scenario for real observations  is that of a
 contaminant star that not centered in the fiber.
{In this case, we cannot consider that both target and contaminant fluxes will completely enter the fiber,
 and therefore we calculated how much of each flux enters the fiber.}
The fraction of binary and fortuitous alignments and their properties also follows established  probability laws. 
To take these parameters into account in a realistic way, 
 we ran Monte Carlo simulations with $10000$ trials, to calculate the distribution function of the impact of the presence 
of a contaminant star on the RV calculation.
Once again, we had to consider both cases of contamination: binaries and fortuitous alignment. If in our simulations one star was
in a binary system and had a background/foreground at the same time, we only considered the stronger RV effect. In these simulations
we  did not take into account the contamination by the sky brightness either, and also just considered that only $\Delta RV \le 20 Km/s$
will produce a significant impact. We ran the simulations for each case of spectral type target considered in previous sections, at 
a distance $d = 1,\,5, \, 10,\,50, \,\mathrm{and} \, 100 \, [pc]$ from Earth.

 Before presenting our results of the statistical analysis we describe in more detail  the considerations to statistically analyze
 the  real binaries (Sect. \ref{BinSA}) and fortuitous alignments (Sect. \ref{FortSA}).

\subsubsection{Statistical analysis of binaries}\label{BinSA}
For our statistical analysis of contamination in binary systems, we considered the mass ratio distributions, 
$q$, mentioned in Sect. \ref{Bin1}.
{We used Table \ref{tab:P|B} based on the study of \cite{Duquennoy_Mayor_1991} for target stars of spectral 
types G and K, and Table \ref{tab:PqM} based on the study of \cite{Janson_2012} for target stars of spectral 
type M.}
 Because we had M1 and M3 spectra for M stars, we used the M1 spectra for [M0, M4[ stars, and for the later types we used the M3 spectrum. 
{From the work of 
\cite{Duquennoy_Mayor_1991} and \cite{Janson_2012}  we also used the probability density function for the orbital period 
 and for the eccentricities (see Table \ref{tab:Pecc}) for G dwarfs, and the
probability density function for the semi-major axis for M-dwarfs (see Table \ref{tab:Pa}).
}
The projected distance $R$ and the $\Delta RV$ were drawn performing a Monte Carlo simulation of Kepler's laws and its projection,
 considering a uniformly random mean anomaly ($M\in [0, 2\pi])$, inclination ($\cos i \in [-1,1]$), 
longitude of the node ($\Omega \in [0, 2\pi]$), and argument of the periapsis ($\omega \in [0, 2\pi] $).

To compute the luminosity of the stellar companion given the mass of the star, we used relation \ref{eq:L_M} from \cite{Duric_2003}:
\begin{equation}\label{eq:L_M} 
\left\{ \begin{array}{ll}
         \frac{L}{L_\odot}\propto\left(\frac{M}{M_\odot}\right)^{4.0} & \mbox{if $M>0.43 M_\odot$};\\
        \frac{L}{L_\odot}\propto 0.23 \left(\frac{M}{M_\odot}\right)^{2.3}, & \mbox{if $M<0.43 M_\odot$}.\end{array}\right.
\end{equation}

\subsubsection{Statistical analysis of fortuitous alignments}\label{FortSA}
For the fortuitous alignment we are only interested in contaminant stars whose flux can enter the fiber. With  a seeing 
 of 0.93, which is the mean value for the La Silla observatory, we measured 
that the fraction of flux that enters the fiber of a star at 2.6 arcsec from the fiber center is  $2\times10^{-8}$.
If target and contaminant star have the same magnitude, this corresponds to an effective  $\Delta m $ of 19 within the fiber, 
and so its impact on the target RV can be considered null for our purpose. We then created a probability density function to have
 a contaminant star of a certain spectral type and magnitude within a radius of  up to 2.6 arcsec around the target star,
using the density tables from Appendix \ref{tables}.

For background/foreground stars within the 2.6 arcsec radius we assumed a uniform distribution, and  if a contaminant is present, 
the probability that the star is standing at a distance $r$ from the target is proportional to $r^2$, and the probability to have 
the fortuitous contaminant star between $r$ and $r+dr$  is
given by equation \ref{eq:pR}:
\begin{equation}
 P(r,r+dr) = \pi(2r dr + dr^2) \label{eq:pR}, 
\end{equation}
in which the members are self-explanatory.

To calculate $\Delta RV$ we used the probability density function for the systemic RV, which was computed by fitting a Gaussian  
 to the systemic velocity of stars from HARPS. This yielded an average and a dispersion of $(\mu,\sigma)=(7.87, 29.87)$.

\subsubsection{Statistical analysis results}
\begin{figure*}[]
\begin{center}
\mbox{
\subfigure[]{\resizebox{0.45\textwidth}{!}{\includegraphics{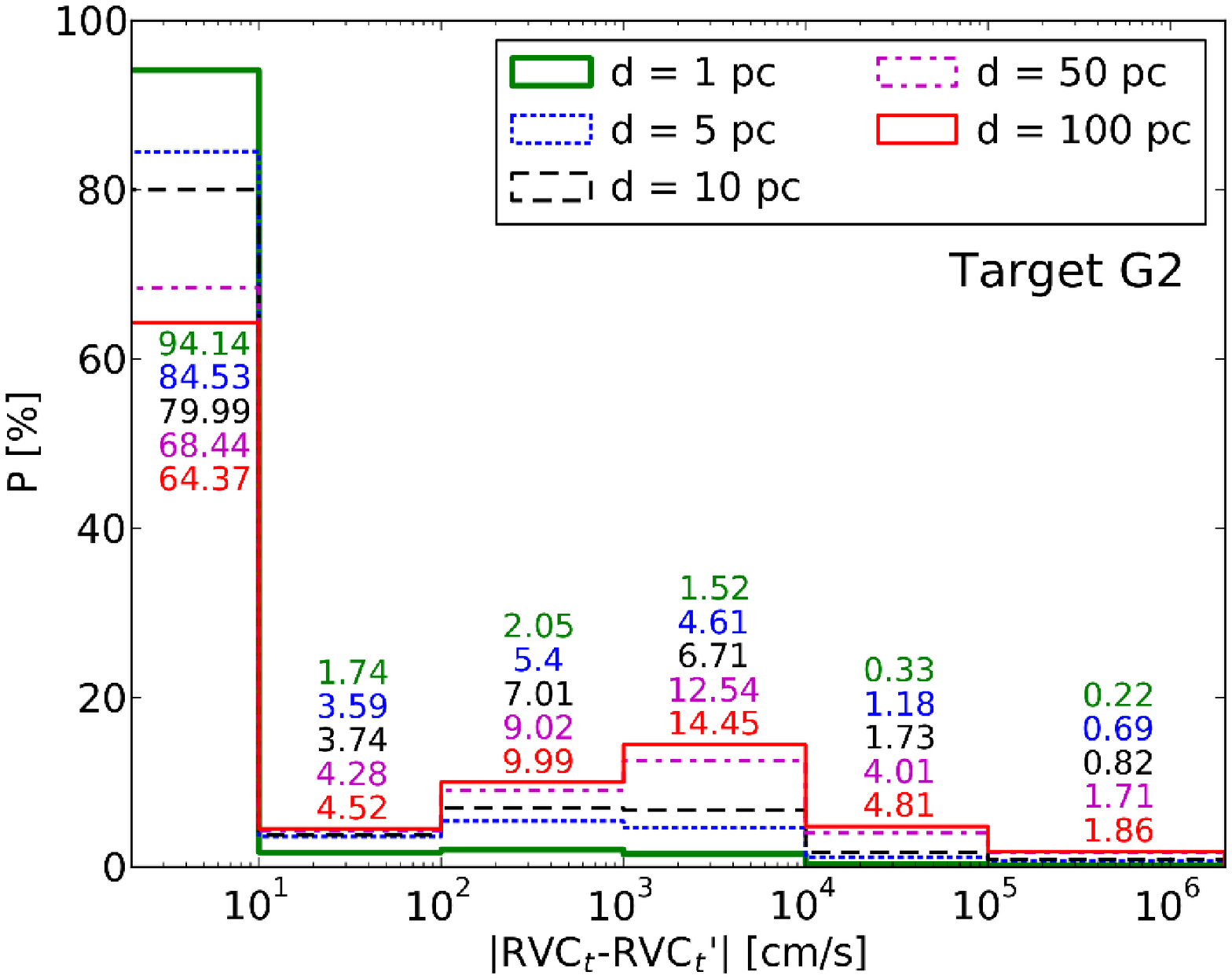}\label{fig:ErroMass1}}}\hspace{0.00cm}}
\mbox{\hspace{0.0cm}
\subfigure[]{\resizebox{0.45\textwidth}{!}{\includegraphics{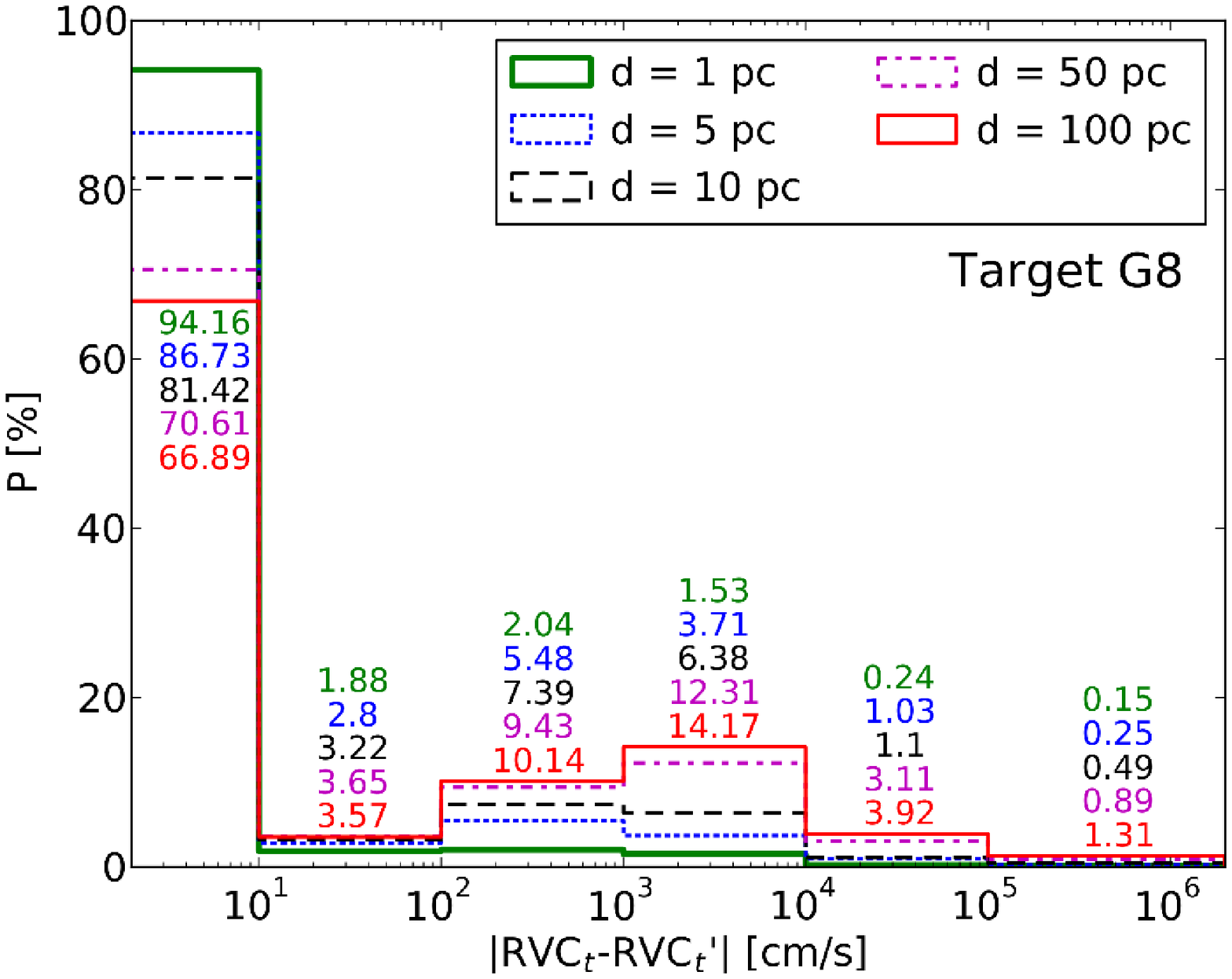}\label{fig:ErroMass09}}}\hspace{0.00cm}}
\mbox{\hspace{0.0cm}
\subfigure[]{\resizebox{0.45\textwidth}{!}{\includegraphics{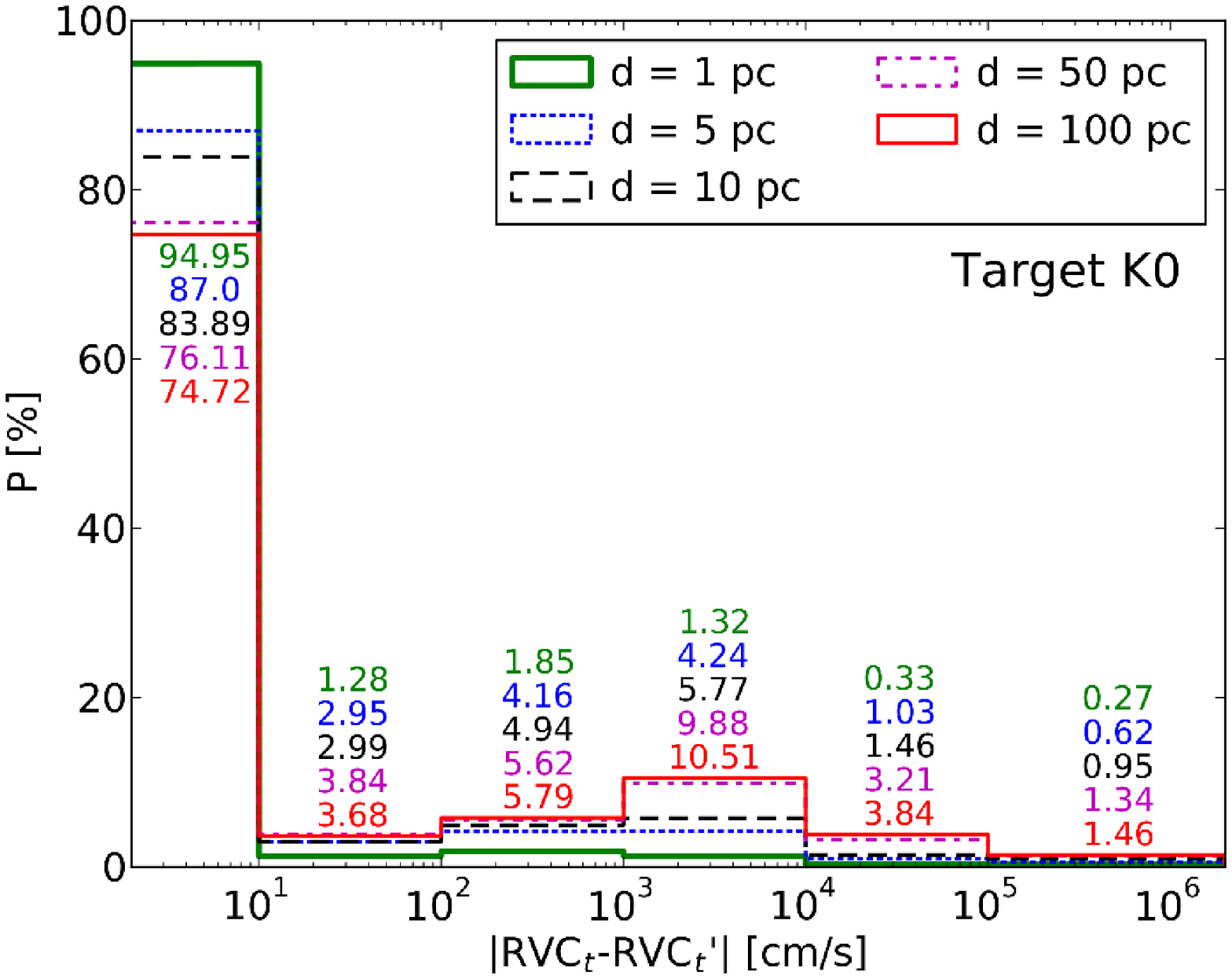}\label{fig:ErroMass07}}}\hspace{0.00cm}}
\mbox{\hspace{0.0cm}
\subfigure[]{\resizebox{0.45\textwidth}{!}{\includegraphics{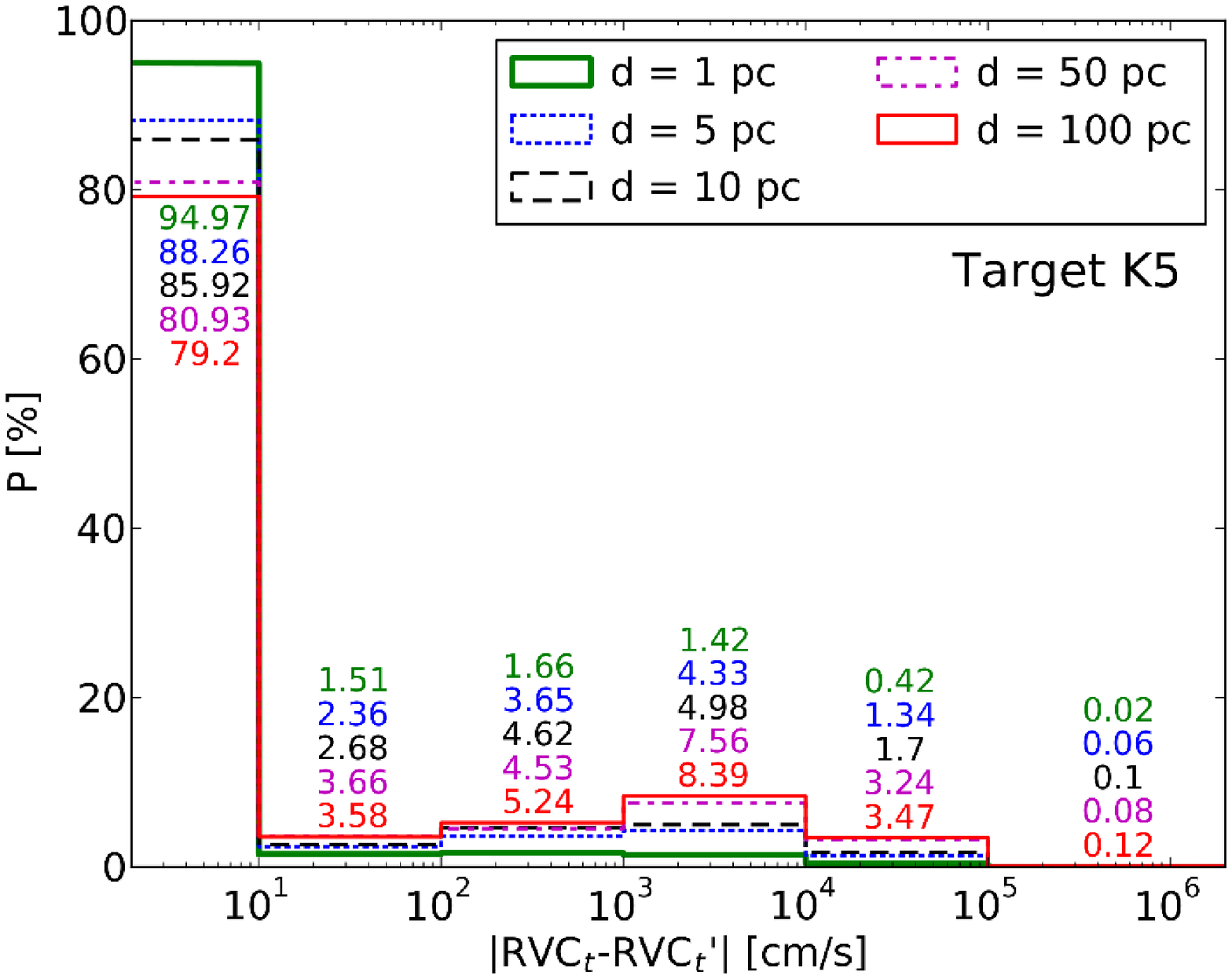}\label{fig:ErroMass06}}}\hspace{0.00cm}}
\mbox{\hspace{0.0cm}
\subfigure[]{\resizebox{0.45\textwidth}{!}{\includegraphics{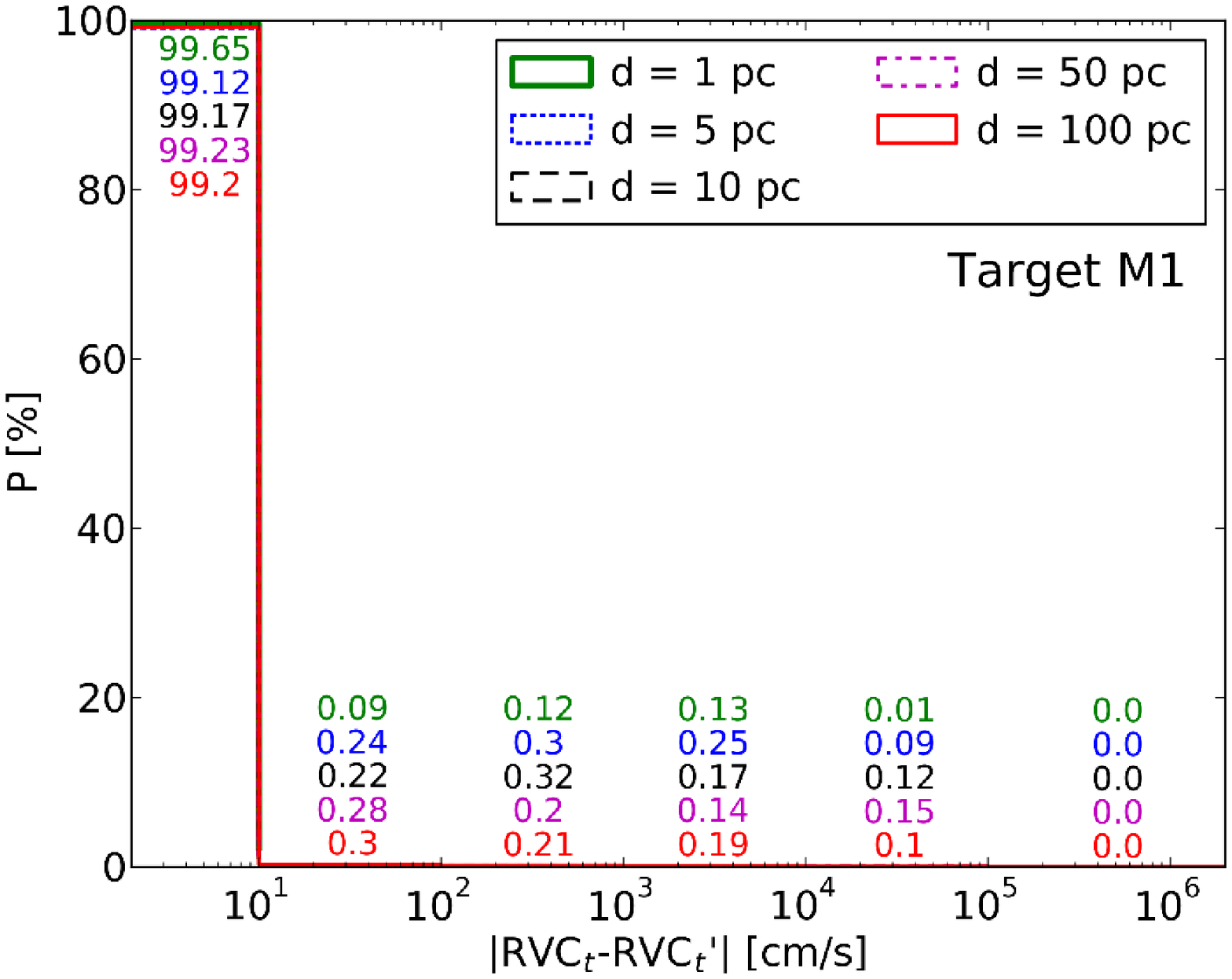}\label{fig:ErroMass05}}}\hspace{0.00cm}}
\mbox{\hspace{0.0cm}
\subfigure[]{\resizebox{0.45\textwidth}{!}{\includegraphics{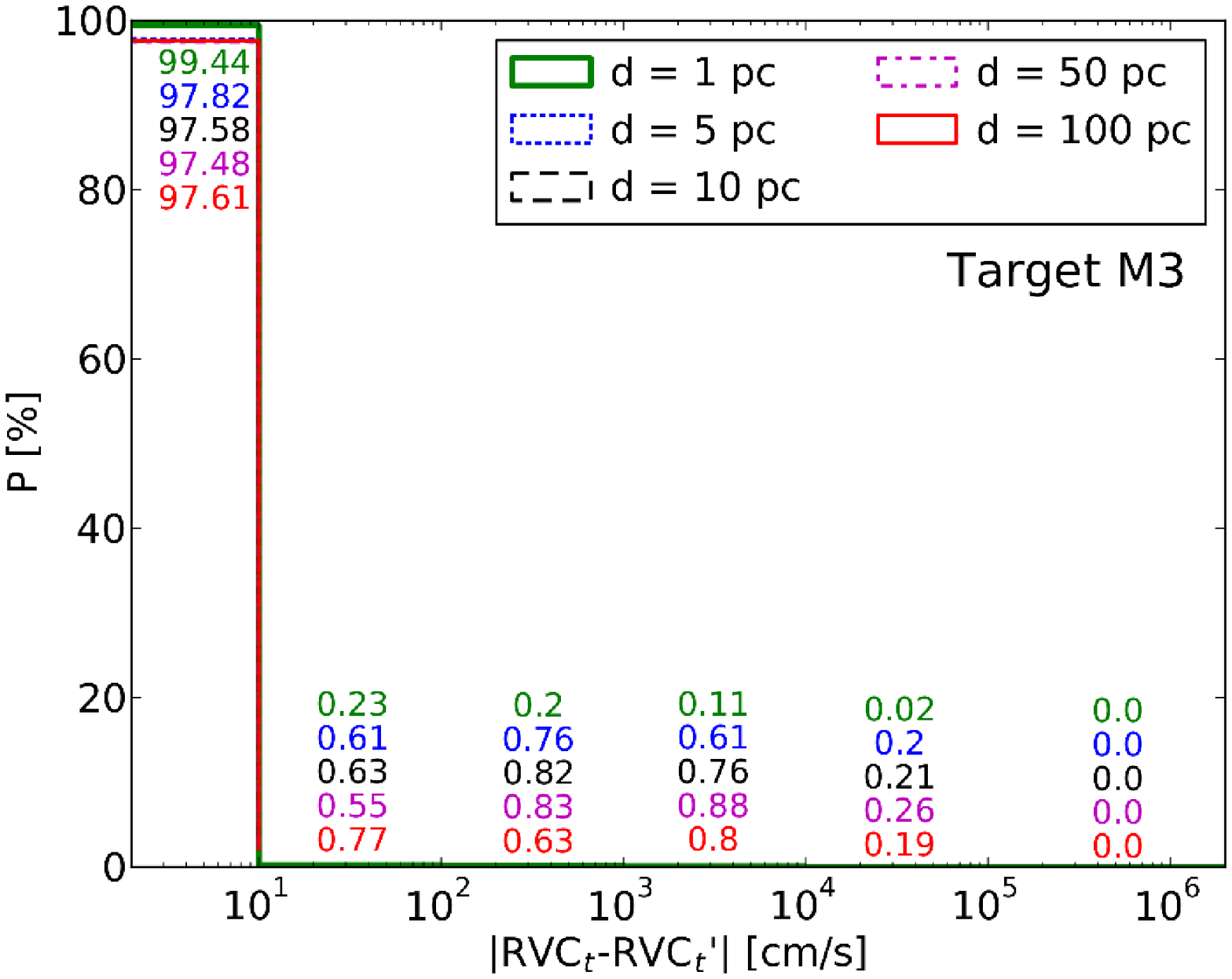}\label{fig:ErroMass02}}}}
\caption{\label{fig:Errocomum} Close-up  of the distribution of the expected contamination , $|RVC_t -RVC_t'|$, 
 on the radial velocity, $|RVC_t|$, of a target star of spectral type \textbf{a)} G2, \textbf{b) }G8,
 \textbf{c) }K0, \textbf{d)} K5, \textbf{e)} M1, and \textbf{f) }M3, and at distance d, caused by a background 
or a companion star. The green thick solid line shows the distribution if the target star stands at a distance of  d = 1 pc, the blue dotted line for d  = 5 pc,
black dashed line for d = 10 pc, magenta dash-dotted line for d = 50 pc, and red solid line for 100 pc. Numbers in each bin correspond to the impact probability
for each target distance: the number in the top corresponds to 1 pc and the number in the bottom to 100 pc.}
\end{center}
\end{figure*}

The probability to have a given impact on the RV calculation for each case is presented in Fig. \ref{fig:Errocomum}.
We can see that the most probable scenario is to have a contamination of less than 10 cm/s.
But contamination may happen and  is more probable for far-away 
target stars, either because they become fainter, or because, in the case of a binary star, it is more probable that the companion star 
flux enters the fiber.
We can examine the target star G8, for instance. In this case, if the target star is 100 pc away, there is only a probability of 67\% for 
a contamination below 10 cm/s, and 14\%  probability for a contamination between $10 $ and $100$ m/s. 
There is a peak around a contamination of $10^3$ cm/s, which is, very probably, caused by physical binary contamination. 
This level of contamination needs to be taken into account when choosing targets for planet searches. This becomes even more important 
when choosing target stars to be observed with ESPRESSO.  

\section{Discussion}\label{sec:Disc}

We showed that we should not neglect the possibility that a stellar companion contaminates the RV 
calculations. This is specially true for far-away stars,  which have a higher probability to have two stars within the fiber.
Nevertheless, we  stress some assumptions we made that may have some impact on our work. Because spectra  with high SN 
are scarce for late-M stars, 
in  our statistical analysis we used the M1 spectra to represent the [M0, M5[ bin and the M3 spectra for the [M5, M9]. We also assumed that the 
star will have at most one stellar companion, because we found that the probability to have two contaminant stars is too low. 
We also did not take into account the effect of the sky brightness if there was no stellar companion.

We  also stress  that the Besan\c{c}on model is a probabilistic model, and thus is not free of errors. We  consider it reasonable
 that the distribution of fortuitous  alignments is a Poisson distribution of stars. Thus, the associated error is equal to $\sqrt{N}$, 
where $N$ is the number of stars per area element.  

The results presented in Fig. \ref{fig:comum} may lead one to think that contamination may be stronger than it actually is, but we should not
forget that this figure only shows the maximum impact of contamination as a function of $\Delta m$, which occurs when 
$|RV(contaminant)|\sim0.6\times FWHM(target)$. We also considered that both stars are well-centered in the fiber and that all  flux enters the
fiber. This justifies our statistical analysis for the impact of the stellar companion within the fiber. With our statistical analysis we are able to conclude 
that it is important to take into account the possibility of contamination by a stellar companion within the fiber, especially when the RV precision
enters the domain of the $cm/s$. Although the probability to have a contamination lower than 10 cm/s is above  50\% for every target star considered, 
we should pay special attention to far-away target stars, for which the impact of contamination can be higher than $10 cm/s$ for  $30 - 40\% $ of the cases.

\subsection{Impact on typical cases}

As a  consequence  of the results presented in Fig.\ref{fig:comum}, we present
 in Table \ref{tbl:impact} a study on the limit $\Delta m$ for which the
  maximum impact on the RV calculation is higher than 10 cm/s, 1, 10, and
    100 m/s. 
    From this table we can see that for a G star with a contaminant M star,
  an impact higher than 100 m/s only occurs if $\Delta m < 2$. However, in Table
   \ref{tbl:moreprob} from Sect. \ref{AA1} we have seen that the probability to have a contaminant M star of $m_v < 17$ is very low. Thus, if we 
are only interested in target stars brighter than $m_v= 15$, an impact higher than 100 m/s is very unlikely. We can also see that when observing
 an M star, an F star can never cause an impact higher than 100 m/s, and an impact of tens m/s is only possible if target and contaminant 
have almost the same magnitude ($\Delta m \lesssim 1.5$). 
On the other hand, even for actual precisions, we may reach a potentially detectable contamination due to a $\Delta m =9$. 
In many cases the induced effect is not detectable
because it is a constant effect. Nevertheless, there are conditions (variable seeing,  centering and/or pointing problems, etc.) in which the induced effect
will not be constant. Even so, a very particular configuration is necessary  (and is highly unlikely) for the effect to create a periodic signal.     
A contamination stronger than 1 m/s on the RV  calculation of a K0 target star is possible if $\Delta m < 9$. If the K0 star is at 10 pc, it will have 
 a $m_v\sim6$, 
which means that contaminant stars with $6\le_v\le15$ can cause such a contamination. The probability to have one contaminant star with $6\le m_v\le15$ within the 0.5'' 
radius fiber is 0.02\% only.

These tabled effects correspond to the maximum case, i.e., to the difference between contamination and no contamination.
Discussed cases are extreme in the sense that they are contamination and no contamination. Nevertheless, to  extensively analyze a
 variable contamination impact we would need to perform many hypotheses with uncertain parameters, such as seeing or flux counts.

\begin {table}[]
  \small
  \centering
  \caption{Limit $\Delta m  $ for which the maximum impact (I$_{max}$)  on the target RV is higher than 0.1, 1, 10, and 100 m/s, for   
each combination of target spctral type (TST) and contaminant spectral type (CST). } \label{tbl:impact}
  \tabcolsep 3.2pt
 \begin{tabular}{c c c c c c c c c c}

\cline{3-10}
& & \multicolumn {8}{ c }{ \textbf{CST}}\\\hline
{\textbf{TST} }& \textbf{I$_{max}$}[m/s] &  F2  &  F5  &  G2  &  G8  &  K0  &  K5  &  M1  & M3   \\\hline
G2             & 0.1              &  10  &  10  &  11  &  11  &  11  &  10  &  9.5 & 9.5  \\
               & 1                &  8   &  8   &  9   &  9   &   9  &  8   &  7   & 7    \\
               & 10               &  5.5 &  5.5 &  6   &  6   &   6  &  5.5 &  4.5 & 4.5  \\
               & 100              &  3   &  3   &  4   &  4   &   4  &  3   &  2   & 2    \\
G8             & 0.1              &  10  &  10  &  11  &  11  &  11  &  10.5&  9.5 & 9.5  \\
               & 1                &  7.5 & 7.5  &  8.5 &  8.5 & 8.5  &  8   &  7   & 7    \\
               & 10               &  5   &  5   &  6   &  6   &   6  &  5.5 &  4.5 & 4.5  \\
               & 100              &  2.5 &  2.5 &  3.5 & 3.5  &  3.5 &  3   &  2   & 2    \\
K0             & 0.1              &  9.5 &  9.5 &  11  &  11  &  11  &  10.5&  9.5 & 9.5  \\
               & 1                &  7.5 & 7.5  &  8.5 &  8.5 & 8.5  &  8   &  7.5 & 7.5  \\
               & 10               &  4.5 &  5   &  6   &  6   &   6  &  5.5 &  4.5 & 4.5  \\
               & 100              &  2.5 &  2   &  3.5 & 3.5  &  3.5 &  3   &  2   & 2    \\
K5             & 0.1              &  10  & 10.5 &  11  &  11.5&  11.5&  11  &  10  & 10   \\
               & 1                &  7.5 & 8    &  8.5 &  9   &  9   &  8.5 &  7.5 & 7.5  \\
               & 10               &  5   &  5.5 &  6   &  6.5 &   6.5&  6   &    5 &   5  \\
               & 100              &  2.5 &  2.5 &  3.5 & 4    &  4   &  3.5 &  2.5 & 2.5  \\
M1             & 0.1              &  6.5 & 6.5  &  7.5 &  8   &  8   &  8.5 &  10.5& 10.5 \\
               & 1                &  4   & 4    &  5.5 &  6   &  5.5 &  6.5 &  8   & 8    \\
               & 10               &  1.5 &  1.5 &  3   &  3   &   3  &  4   &  5.5 &  5.5 \\
               & 100              & ---  &  --- &  0.5 & 0.5  &  0.5 &  1.5 &  2.5 & 2.5  \\
M3             & 0.1              &  6.5 & 6.5  &  7.5 &  8   &  8   &  9   &  10.5& 10.5 \\
               & 1                &  3.5 & 3.5  &  5   &  5.5 &  5.5 &  6.5 &  7.5 & 7.5  \\
               & 10               &  1   &  1   &  2.5 &  3   &   3  &  3.5 &  5   &  5   \\
               & 100              & ---  &  --- &  0   & 0.5  &  0.5 &  1.5 &  2.5 & 2.5  \\
\hline
 \end{tabular}
 \end{table}

\subsubsection{Sky brightness}
In our analysis of the sky brightness impact  we assumed that sky brightness is mainly caused by the moon, and so its spectra will be that 
of a star with spectral type G2. The new-moon sky brightness should then be considered as a lower limit for the contamination from the sky
 brightness with a G2 spectrum. However, we should not forget that sky brightness has other sources and may depend on the sky zone. Sky spectra 
from the Galactic disk are surely different from the Galactic halo. Moreover, there is also the possibility to have a galaxy as contaminant, 
which was not considered. Nevertheless, we consider it a good approximation, or at least a good starting point, to treat the sky 
brightness as a G2 star.

If observations are carried out on a night with moon, and if cirrus are present, the contamination by the moon will be enhanced.

{ \citet {Tripathi_2010}, reported two
 outlier RV data points caused by cirrus during their observations of WASP-3,  an F7 star of $m_v=10.7$,  on a full-moon night.
 Their RV measurements present a redshift of 140 m/s 
and of 49 m/s  with respect to their best-fitting models. 
If we assume that cirrus will reflect some of the moonlight,  we  have a G2 contaminant. 
The maximum value considered for all practical effects for the  visual magnitude of this contaminant will be that of
 the target star (if it were higher than the target magnitude, 
the target star would be no longer visible).
 Thus, for a G2 target star with $m_v =10.7$ and a contaminant with the same spectral type and magnitude, the 
maximum contamination will be 2.7 Km/s.
This corresponds to 0.15\% of the moonlight flux, considering that 100\% of the moonlight flux corresponds to a star of magnitude 3.65. 
If 0.01\% of the moonlight reaches the fiber, corresponding to a contaminant of spectral type G2 with $m_v=13.7$, the contamination would be 140 m/s.
For a  contamination of 49 m/s, 0.003\% of the moonlight flux would be enough, corresponding to a G2 contaminant with $m_v=14.8$.
}
Because the \citet {Tripathi_2010} data were taken with the High Resolution Echelle Spectrometer (HIRES) on the Keck I 
telescope, and we used HARPS spectra in our simulation, {and also because we did not calculate the impact for a target of spectral type F}, 
these contamination values are presented here just as a reference, since we cannot directly compare them. 
{Nevertheless, we consider that their hypothesis of cirrus contamination is valid.}
 
{The sky brightness contamination is not easily removed.
 Although its signal may be strong enough to contaminate the RV calculations, 
it will be too weak to be adequately characterized, fitted, and removed, even for a long-slit spectrograph.}

\subsubsection{Kepler stars}
Our work can also be a valuable asset for the follow-up of programs such as Kepler. 
Kepler stars have $11 \lesssim m_v \lesssim 18$  \citep[see][]{Batalha_2012}, 
and consequently,  measuring the RV becomes challenging. Beyond the problem of these being faint stars, there is also the problem of the false-positive 
planet transit. 
As an example, we consider the case of Kepler-14b \citep{Buchhave_2011}, a planet orbiting an F star with a companion star of nearly the same
magnitude and only 0.3 arcesec of sky-projected angular separation. Buchhave and collaborators were not able to detect a second peak on the CCF
 and  concluded that the two stars probably
have nearly the same RV. They computed an impact of 280 m/s on the RV caused by the companion star. 
A direct comparison with their work is not possible, because their observations were carried out with the
 Fiber-fed \'{E}chelle Spectrograph (FIES) at the 2.5m NOT at La Palma, and with the High Resolution Echelle Spectrometer (HIRES) 
mounted on the Keck I on Mauna Kea, Hawaii.
Moreover, our simulation only contemplates target stars of spectral GKM, and this is an F star.
Still, performing a rough comparison, we see that our simulations indicate an impact of
 $\sim 276$ m/s for a G2 target star with a contaminant of the same spectral type,
 $\Delta m = 1$, and $\Delta RV = 1$.
      

\section{Conclusions}\label{sec:Concl}
{This study on the impact of stellar companions within the fiber of  RV calculations allows us to conclude 
that we should not neglect the possibility of contaminant flux from stellar companions stars, specially 
if we are observing far-away stars. 
On average, if we are observing a G or a K star, the contamination may be higher than 10 cm/s if the difference between target and 
contaminant magnitude is  $\Delta m < 10$, and higher than 1 m/s if $\Delta m < 8$. 
If the target star is a star of spectral type M, 
 a $\Delta m < 8$ is enough   
to obtain the same contamination of 10 cm/s, and a contamination may be higher than 1 m/s  if $\Delta m < 6$.
We also showed that  sky brightness should not be discarded, particularly on full-moon nights, 
if one observes faint target stars ($m_v > 10$). 

These results will allow us to more wisely choose target stars to be observed with 
instruments such as ESPRESSO and CODEX, 
and they provide reference values for the different 
cases of contamination possible in fiber-fed high-resolution spectrographs.
We  also stress that there are diagnosis methods that should be capable of detecting most of the blends that mimick planets.
 We can detect a blend through 
bisector analysis of the CCF or from correlations using different templates \citep{Santos_2002}.      }
\begin{acknowledgements}
{This research has made use of the SIMBAD database,
operated at CDS, Strasbourg, France.
DC, PF, NCS, and GB acknowledge the support from  the European Research 
Council/European Community under the FP7 through Starting Grant agreement number 239953,
 as well as from Funda\c{c}\~ao para a Ci\^encia e a Tecnologia (FCT), Portugal, in the form of grant
reference PTDC/CTE-AST/098528/2008.
We also thank  J. S. Amaral for a critical reading of the manuscipt.}

\end{acknowledgements}


\bibliographystyle{aa}
\bibliography{Diana}

\newpage
\onecolumn
\appendix 
\section{Density of stars}\label{tables}
\begin{figure*}[htbp]
  \centering
  \includegraphics[width=\textwidth, angle=0]{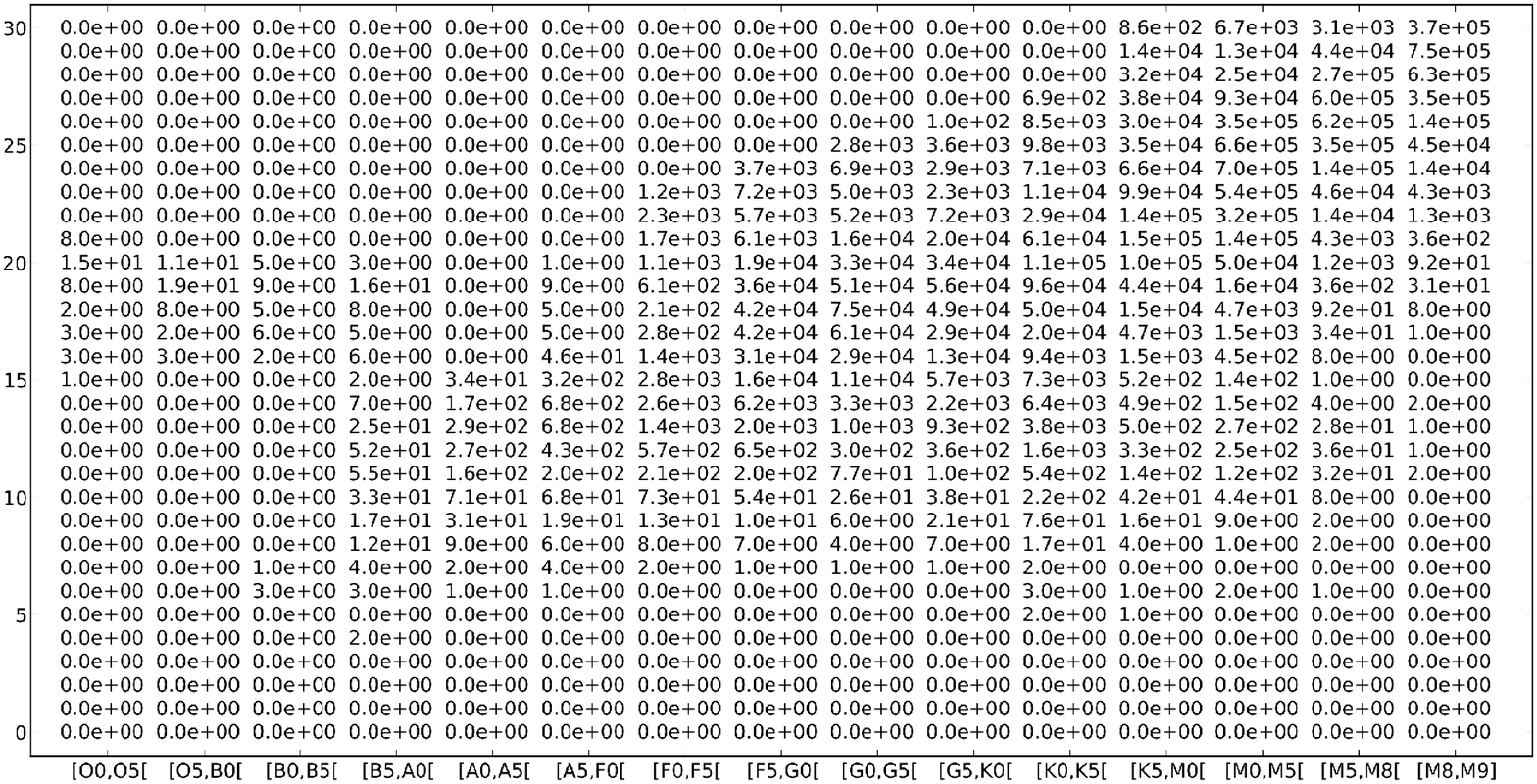}
  \caption{Number of stars in each bin of spectral type and magnitude for an area of the sky with a radius of $3^{\circ}$.}
  \label{tbl:Ntot_Probs_Modele_3grausraio}
\end{figure*}
  
\begin{figure*}[htbp]
  \centering
  \includegraphics[width=\textwidth, angle=0]{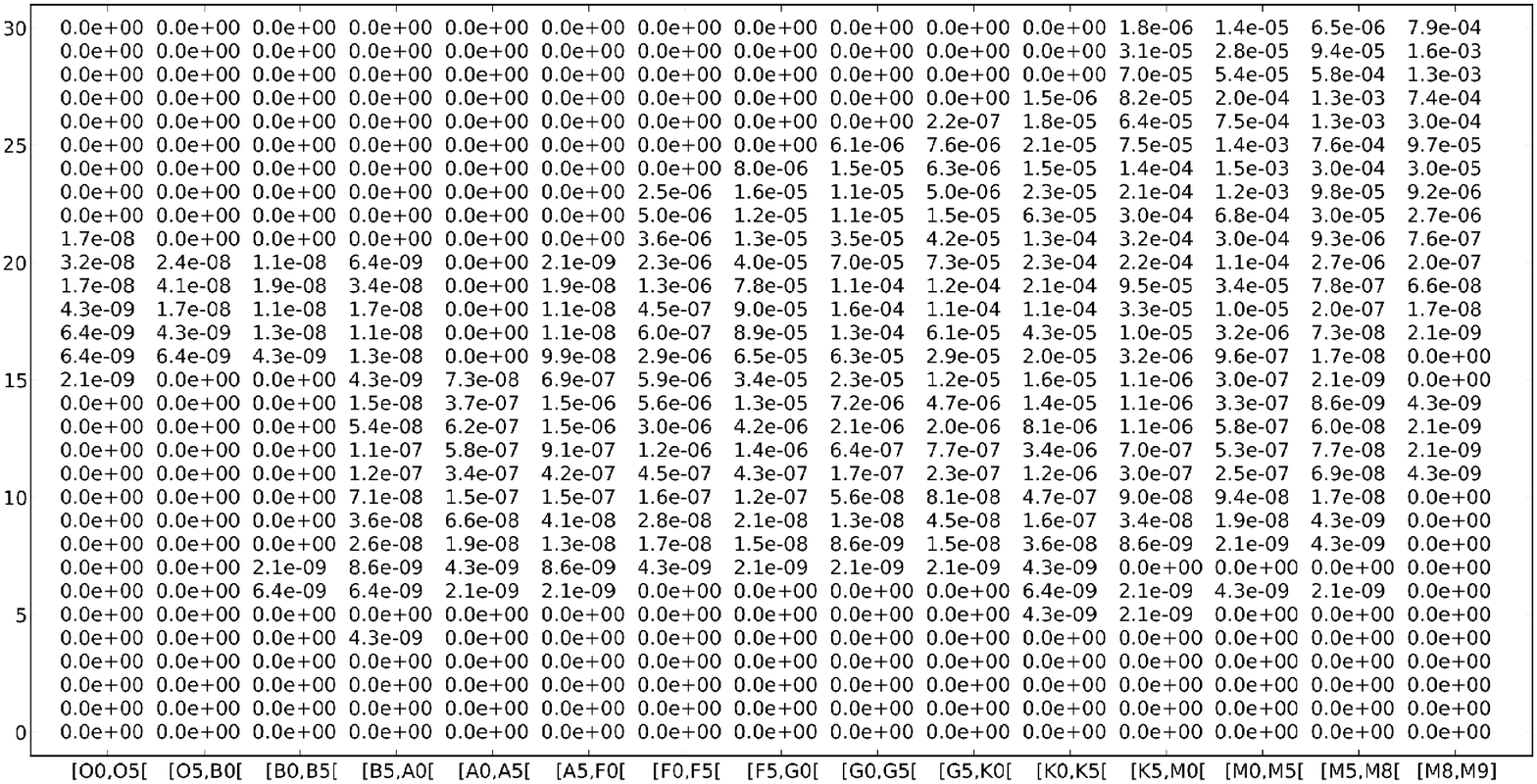}
  \caption{Density of stars in each bin of spectral type and magnitude for an area of the sky with a radius of $0.5$ arcsec.}
  \label{tbl:Probs_Modele_3grausraio}
\end{figure*}
\onecolumn
\section{Method details}\label{details}
To  control  the magnitude of the contaminant star we first  used the relation \ref{eq:m} to calculate $m_v$ 
in the target reference frame,
\begin{equation}\label{eq:m}
  m_t-m_c=-2.5 \times \log (F_t/F_c)\,,
 \end{equation}
where $F_t$ and $F_c$ are the fluxes of the target and of the contaminant.
 The magnitudes are now calibrated and can be changed by multiplying the flux by a factor of 
\begin{equation}
 f=10^{-x/2.5},
\end{equation}
where $x$ is the difference between the new chosen magnitude and the true magnitude of the star
 in the target-star system.

To change the RV of the contaminant spectra we also had to modify them. 
The first challenge  when processing our co-added spectra with the HARPS RV 
pipeline is that it will take into account the header (and proprieties) of the parent spectra. 
To compensate for this  we first need to replace the header of the contaminant spectrum file with that from the target, and then run the cross
correlation function (CCF) recipe of the RV pipeline.   
The next step in this  process was to shift the contaminant spectra. To assess by how much the contaminant 
spectra had to be shifted, we used  equation \ref{eq:wl}
\begin{equation}\label{eq:wl}
 \lambda_{c}=\lambda_{t}\times\left(1+\left(\frac{RVC_c -RVC_t - \Delta RV }{c}\right)\right) \,,
\end{equation}
 where $\lambda_{c}$ are the new wavelengths to derive the desired $\Delta RV$ 
(difference between the RV of the target and the contaminant), $\lambda_{t}$ are the wavelengths
 from the wave file corresponding to the target, and $c$ is the light speed in  vacuum.
These new wavelengths were used to calculate, by  linear interpolation, the number of pixels by which the 
contaminant needed to be shifted. This gives a new $xx$ axis that it was used, recurring to another linear
 interpolation, to shift the contaminant spectra. With the target RV for the contaminant, the next step is
 the sum of the target with the modified contaminant spectra, which is the sum of their fluxes.
The shift of the contaminant spectra should not be larger that the  maximum 
barycentric Earth radial velocity (BERVMX) 
of the star, otherwise there is the risk that the spectral lines enter and leave the orders when performing the CCF calculation.  BERVMX defines the 
excluded spectral lines from the correlation due to the orbital movement of the Earth toward the star. Our goal is to
 introduce a shift equivalent to a BERV value lower than the $\mid$BERVMX$\mid$, so that we do not need to exclude new lines.   

\end{document}